\documentclass[%
preprint,
superscriptaddress,
preprintnumbers,
nofootinbib,
amsmath,amssymb,
aps,
prd,
]{revtex4-1}
\usepackage[utf8]{inputenc}
\usepackage{bm,amsfonts}
\usepackage{graphicx}
\usepackage{subfigure}
\usepackage{hyperref}
\usepackage{color}

\begin{document}

\title{Effective reflected entropy and entanglement negativity for general 2D eternal black holes}
\author{Jiong Lin}
\email{jionglin@hust.edu.cn}
\affiliation{School of Physics, Huazhong University of Science and Technology, Wuhan, Hubei
430074, China}

\author{Yizhou Lu}
\email{Corresponding author.louischou@hust.edu.cn}
\affiliation{School of Physics, Huazhong University of Science and Technology, Wuhan, Hubei
430074, China}

\begin{abstract}
Both reflected entropy and entanglement negativity provide valid measures of entanglement between subsystems of a mixed state. 
For general 2D eternal black holes coupled to CFT matters in large $c$ limit, we perform the replica-trick computation and find that both effective Renyi reflected entropy and effective entanglement negativity can be expressed in terms of the combination of  modified backreacting cosmic branes in $\mathrm{AdS}_3$ bulk.
We then develop a holographic scheme to calculate effective reflected entropy and entanglement negativity for general 2D eternal black holes coupled to CFT matters in large $c$ limit. 
Using the scheme, we check the consistency condition of the island formulae for entanglement negativity and reflected entropy.
We find that the combinations of modified backreacting cosmic branes in $\mathrm{AdS}_3$ bulk from the two island proposals of entanglement negativity exactly match with each other. 
Finally, we study the saturation of the reflected entropy inequality. 

\end{abstract}

\maketitle

\noindent

\tableofcontents

\section{Introduction}
Black holes open a phenomenological window towards quantum gravity. Coupled to a quantum field, black holes emit radiation with the entropy proportional to the area of the horizon. 
For an evaporating black hole formed from matters in a pure state, according to Hawking's calculation, the entanglement entropy of the radiation increases constantly and at Page time, exceeds the thermal entropy of black holes, which is contradictory to unitary evolution in quantum theory, leading to the information loss paradox \cite{Hawking:1974sw,Hawking:1976ra}.
The unitarity requires the entanglement entropy of the radiation follows the Page curve \cite{Page:1993wv,Page:2013dx,Page:1993df}.

Recent progress indicates that when calculating the entanglement entropy of the radiation, we should include part of black hole interior called \emph{islands} \cite{Penington:2019npb,Penington:2019kki,Almheiri:2019psf,Almheiri:2019qdq,Almheiri:2019hni} and the entanglement entropy of the radiation $\mathcal R$ is given by
\begin{equation}\label{island_formula}
S(\mathcal{R})=\mathrm{Min}_{\partial  I}\left\{
\mathrm{Ext}_{\partial I}\left[\frac{\mathrm{Area}(\partial I)}{4G_N}+S_{\mathrm{eff}}({\mathcal{R}\cup I})\right]
\right\},
\end{equation}
where the boundaries of islands are the quantum extremal surfaces (QES) \cite{Ryu:2006bv,Hubeny:2007xt,Engelhardt:2014gca} obtained by extremizing the generalized entropy \cite{Lewkowycz:2013nqa}.
The island formula was studied in the context of 2D Jackiw-Teitelboim (JT) gravity \cite{Almheiri:2014cka} coupled to CFT matters first \cite{Almheiri:2019hni}.
It was then extended to black holes that are not asymptotically AdS, and was even extended to higher dimensions \cite{Almheiri:2019psy,Hashimoto:2020cas,Wang:2021woy,Karananas:2020fwx,Ahn:2021chg,Kim:2021gzd,Ling:2020laa,Alishahiha:2020qza,Anegawa:2020ezn,Geng:2020qvw,Bak:2020enw,Chu:2021gdb,Yu:2021cgi,Lu:2021gmv,Nam:2021bml,He:2021mst,Yu:2021rfg}. 
Islands for dynamically evaporating black holes was considered in Refs. \cite{Goto:2020wnk,Hollowood:2020cou,Gautason:2020tmk,Gan:2022jay}.
The island proposal also triggered the discussions on cosmology \cite{Hartman:2020khs,VanRaamsdonk:2020tlr,Balasubramanian:2020xqf,Sybesma:2020fxg,Fallows:2021sge,Espindola:2022fqb}.
For other relevant comments on island and Page curve, please see Refs. \cite{Geng:2021hlu,Geng:2020fxl,Miyata:2021qsm,Anderson:2020vwi,Chen:2020uac,Chen:2020hmv,Hernandez:2020nem,Krishnan:2020oun,Krishnan:2020fer,Caceres:2020jcn,Li:2021lfo,Bhattacharya:2021dnd,Ghosh:2021axl,Bhattacharya:2021nqj,Iizuka:2021tut,Geng:2021mic,Uhlemann:2021nhu,Uhlemann:2021itz,Bhattacharya:2021jrn,Omidi:2021opl,Kames-King:2021etp,Verheijden:2021yrb,Caceres:2021fuw,Bianchi:2022ulu,Seo:2022ezk,Grimaldi:2022suv,Bousso:2022ntt,Suzuki:2022xwv}.

Though, the island formula \eqref{island_formula} is originally obtained from AdS/CFT holography, it can be derived from bulk gravitational path integral \cite{Almheiri:2019qdq,Penington:2019kki}.
To be specific, one first evaluates the Renyi entropy
\begin{equation}
    S^{(n)}=\frac{1}{1-n}\log \text{Tr}(\rho^n)
\end{equation}
by employing the replica trick \cite{Callan:1994py,Holzhey:1994we,Calabrese:2009qy,Casini:2009sr} and then take the limit $n\to1$ to get the entanglement entropy 
\begin{equation}
    S=\lim_{n\to1} S^{(n)}.
\end{equation}
Note that performing gravity path integral with replica trick involves all possible solutions of spacetime configuration subject to certain boundary conditions. 
In addition to the Hawking disk saddle, we should also consider the contribution from replica wormhole saddle, which connects $n$ different sheets of replica geometry. 
The replica wormhole introduces branch cut in the original spacetime, which corresponds to the boundary of island. 
Before the Page time, the Hawking disk saddle dominates and the  entanglement entropy of the radiation increases. 
After the Page time, the replica wormholes saddle dominates so that the entanglement entropy decreases. 

To further uncover the entanglement structure within the Hawking radiation, it is necessary to resort to some measures of entanglement for mixed states, for example, the entanglement of purification \cite{Takayanagi:2017knl,Nguyen:2017yqw,Liu:2019qje,Ghodrati:2019hnn,BabaeiVelni:2019pkw,Jokela:2019ebz,Bhattacharya:2020ymw} and canonical purification \cite{Dutta:2019gen}. 
\emph{Reflected entropy} \cite{Dutta:2019gen,Bao:2019zqc,Chu:2019etd,Akers:2019gcv,Jeong:2019xdr,Kudler-Flam:2020url,Moosa:2020vcs,Bueno:2020vnx} has been proposed as the von Neumann entropy in a canonically purified state $|\sqrt{\rho_{AB}}\rangle_{ABA^*B^*}$ in the doubled Hilbert space $\left(\mathcal{H}_{A} \otimes \mathcal{H}_{A}^{\star}\right) \otimes\left(\mathcal{H}_{B} \otimes \mathcal{H}_{B}^{\star}\right)$, i.e.
\begin{equation}
    S_R(A:B)=S(AA^*)_{|\sqrt{\rho_{AB}}\rangle},
\end{equation}
which serves as a valid measure of entanglement between $A$ and $B$.
In the classical gravity limit, reflected entropy can be holographically expressed in terms of twice the  minimal entanglement wedge cross section (EWCS) \cite{Dutta:2019gen}. 
For recent comments on EWCS, please see Refs. \cite{BabaeiVelni:2020wfl,Khoeini-Moghaddam:2020ymm,Sahraei:2021wqn}.
Taking quantum correction into account, to $\mathcal{O}(G^0_N)$ and all order of $\hbar$, with the analog of Engelhardt and Wall prescription \cite{Engelhardt:2014gca}, the reflected entropy is conjectured as \cite{Li:2020ceg}
    \begin{equation}
S_{R}(A: B)=\operatorname{ext}_{Q^{\prime}}\left\{\frac{2 \operatorname{Area}\left(Q^{\prime}=\partial a \cap \partial b\right)}{4 G_{N}}+S_{R}^{\text {bulk }}(a: b)\right\},
\end{equation}
where $Q^{\prime}=\partial a \cap \partial b$ is the quantum entanglement wedge cross section (QEWCS). 
In semiclassical gravity, the island formula for reflected entropy has been proposed in Refs. \cite{Chandrasekaran:2020qtn,Li:2020ceg}.
However, the replica-trick computation of bulk effective reflected entropy including island contribution involves multi-point correlation functions of twist operators, which makes the calculation very difficult. 
In our paper, we will develop a holographic scheme to calculate the effective Renyi reflected entropy for general 2D eternal black holes coupled to CFT matters in the large $c$ limit. 

There is another important measure of entanglement for mixed states, the \emph{entanglement negativity} \cite{Vidal:2002zz,Calabrese:2012ew,Calabrese:2012nk,Rangamani:2014ywa,Chaturvedi:2016rcn,Chaturvedi:2017znc,Malvimat:2017yaj,Malvimat:2018izs,Dong:2021clv}, which is an upper bound of distillable entanglement. 
The logarithmic negativity is defined as
\begin{equation}\label{ee}
\mathcal{E}\left(\rho_{A B}\right)=
\log \left\|\rho_{A B}^{T_{B}}\right\|_{1},
\end{equation}
where $\rho_{A B}^{T_{B}}$ is the partially transposed density matrix and $\left\|\ \right\|_{1}$ denotes the Schatten 1-norm. 
Note that entanglement negativity is related to the number of Bell pairs one can extract.
For separable states, the logarithmic negativity \eqref{ee} vanishes.
It has been noted that holographically, the entanglement negativity in $\mathrm{CFT}_2$ between adjacent intervals or disjoint intervals in proximity can be  expressed in terms of a combination of the areas of non-backreacting cosmic branes in $\mathrm{AdS}_3$ \cite{Jain:2017aqk,Jain:2017uhe,Malvimat:2018txq,Malvimat:2018ood}. 
On the other hand, holographic entanglement negativity can be also determined by the area of a backreacting cosmic brane on the entanglement wedge cross section (EWCS) \cite{Kudler-Flam:2018qjo} or by half the Renyi reflected entropy of order 1/2 \cite{Kusuki:2019zsp}.
Note that in the context of $\mathrm{AdS}_3/\mathrm{CFT}_2$, the entanglement negativity from the sum of non-backreacting cosmic branes in $\mathrm{AdS}_3$ precisely matches with that from EWCS. 
In semiclassical graivty, two island formulae for entanglement negativity has been proposed in Ref. \cite{KumarBasak:2020ams}.
One is given by the combination of the Renyi entropy of order 1/2 and the other is given by extremizing the generalized entanglement negativity or half the generalized Renyi reflected entropy of order 1/2. 
It has been shown in Ref. \cite{KumarBasak:2020ams} that two proposals exactly match with each other for JT black holes coupled to a bath. 
Recently, replica wormhole for entanglement negativity has been advanced in Ref. \cite{Dong:2021oad}. 
For other recent comments on entanglement negativity and reflected entropy, see Refs \cite{KumarBasak:2021rrx,Afrasiar:2021hld,Ling:2021vxe,Basak:2022cjs,Setare:2022uid,Basu:2022nds,Akers:2022max,Basak:2022acg}.

Similar to the effective reflected entropy, it is a challenge to perform the replica-trick computation of bulk entanglement negativity to include the contribution from islands, because of the emergence of multi-points correlation functions.
In our paper, we will also develop a holographic calculation scheme for effective entanglement negativity for general 2D eternal black holes coupled to CFT matters in the large $c$ limit. 
With the holographic scheme in hand, we will study the consistency condition for entanglement negativity and reflected entropy.  

The paper is organized as follows.
In Sec.\ref{sec2} and Sec.\ref{sec3},  we will perform the replica-trick computation of effective Renyi reflected entropy and effective entanglement negativity for general 2D eternal black holes coupled to CFT matters in large $c$ limit. 
From that, we will extract the holographic calculation scheme. 
Sec.\ref{sec4} and Sec.\ref{sec5} are devoted to the consistency check for island formulae for entanglement negativity and reflected entropy. 
We will show the exact match of the combinations of modified backreacting cosmic branes from two island proposals of entanglement negativity in Sec.\ref{sec4}. 
In Sec.\ref{sec5}, we study the saturation of reflected entropy inequality in by utilizing our holographic scheme. 
We conclude our paper in Sec. \ref{sec7}.

\section{Effective Renyi reflected entropy}\label{sec2}
In this section, we will propose a holographic scheme for effective Renyi reflected entropy for general 2D eternal black holes coupled to CFT matters in large $c$ limit. 
The metric of 2D eternal black holes can be cast into a conformally-flat form
\begin{equation}
    \mathrm{d}s^2=-\frac{1}{\omega^2}\mathrm{d}y_{R,L}^+\mathrm{d}y_{R,L}^-,
\end{equation}
where $y^{\pm}_{R,L}$ is the thermal coordinate with the inverse temperature $\beta$.
The subscripts $R,L$ denote the right and left Rindler wedge, respectively.
Under conformal transformation 
\begin{equation}\label{trans}
    x_R^{\pm}=\pm e^{\pm2\pi\beta^{-1}y_R^{\pm}},\ 
    x_L^{\pm}=\mp e^{\mp2\pi\beta^{-1}y_L^{\pm}},
\end{equation}
the whole spacetime can be related to vacuum coordinate $x^{\pm}$ and the vacuum metric is 
\begin{equation}
    \mathrm{d}s^2=-\frac{1}{\Omega^2}\mathrm{d}x^+\mathrm{d}x^-.
\end{equation}

\begin{figure}[htp]
\centering
\includegraphics[width=0.8\textwidth]{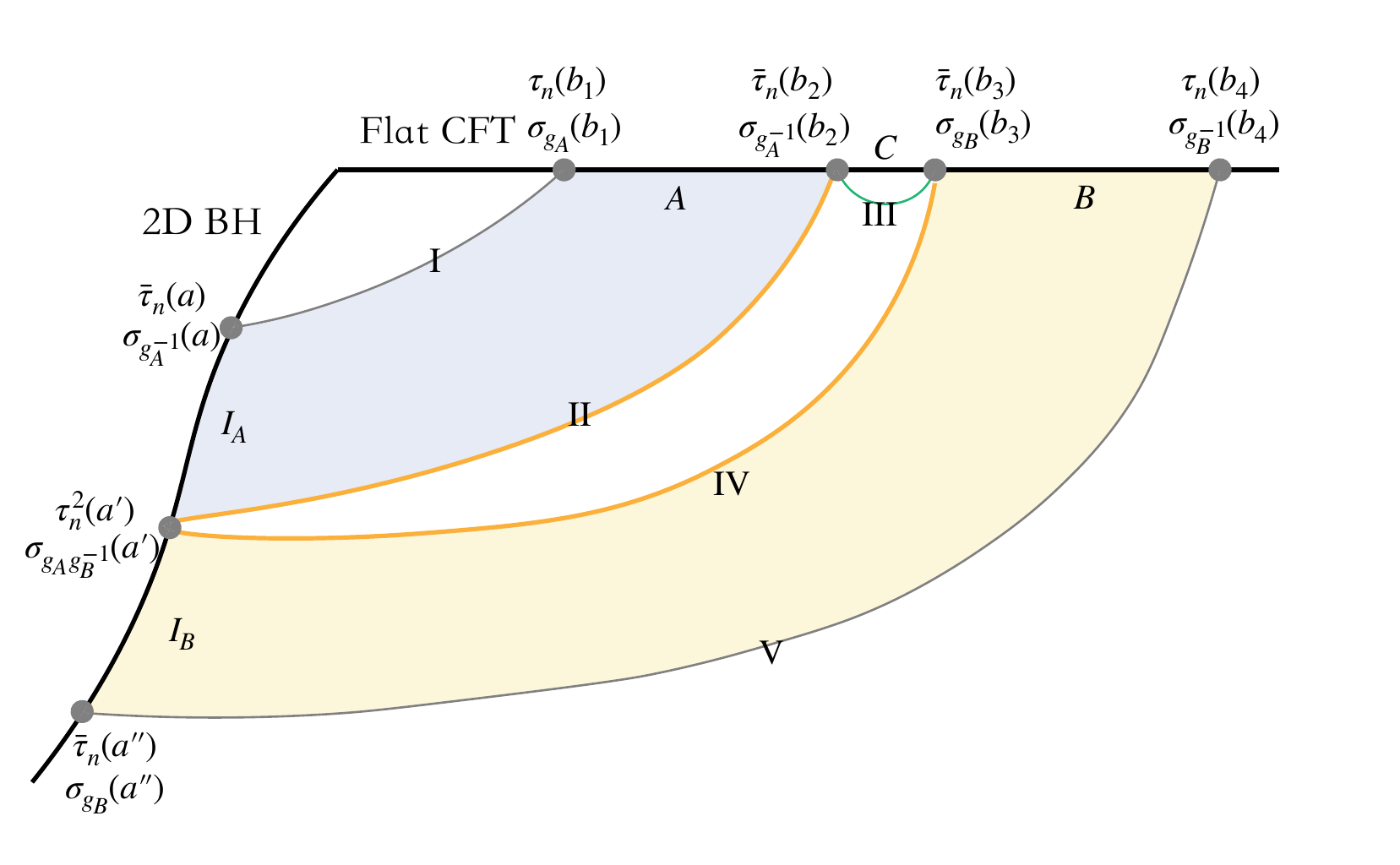}
\caption{Phase-I: A and B are large and admit the islands $I_A$ and $I_B$.
The sandwich interval C is too small to admit an island.
The light blue and light yellow shadows correspond to the entanglement wedge of  $A \cup I_{R}(A)$ and $B \cup I_{R}(B)$, respectively.
$\sigma$ and $\tau$ denote the twist operators for reflected entropy and entanglement negativity, respectively.}
\label{Fig:1}
\end{figure}

\begin{figure}[htp]
\centering
\includegraphics[width=0.47\textwidth]{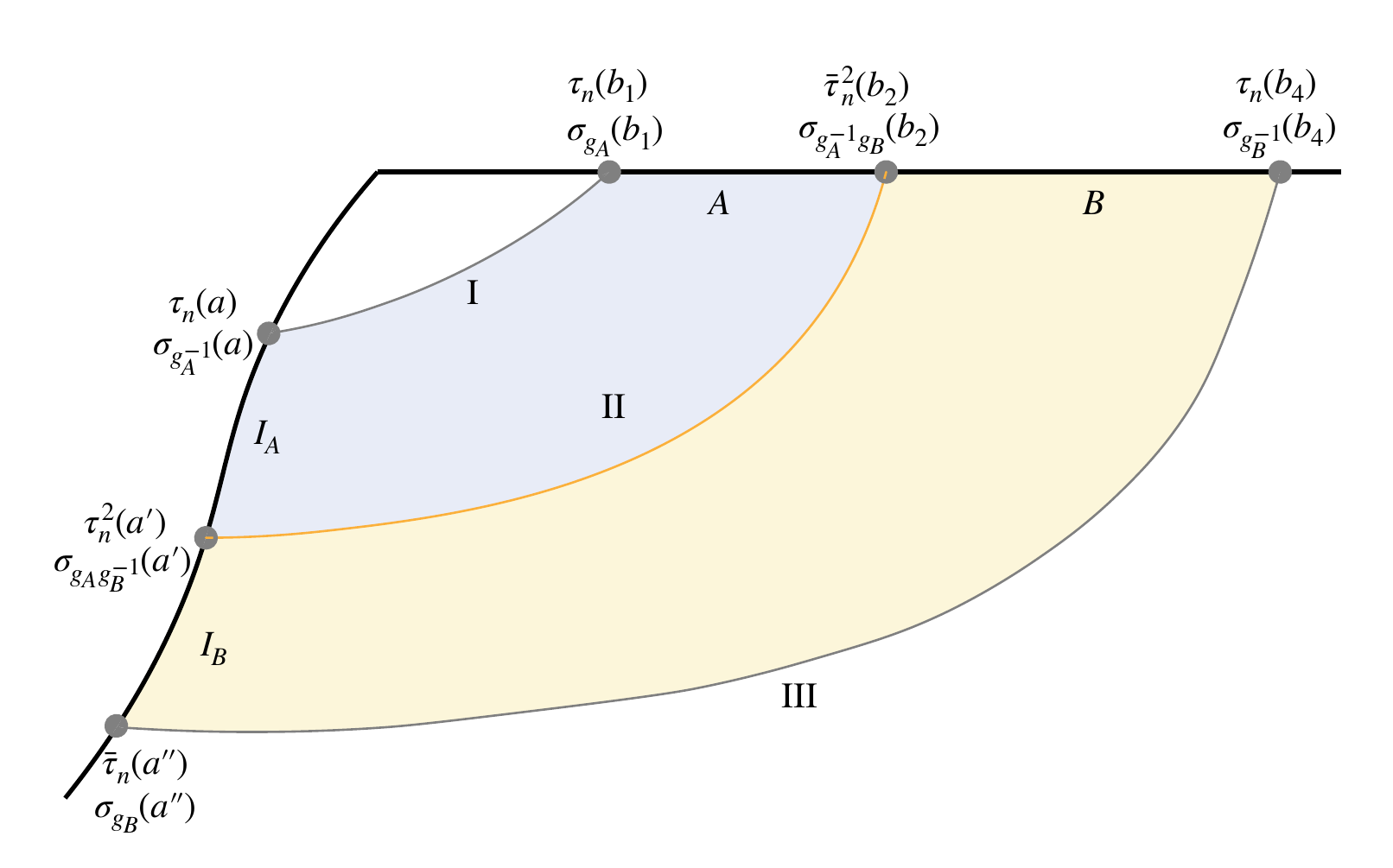}
\includegraphics[width=0.47\textwidth]{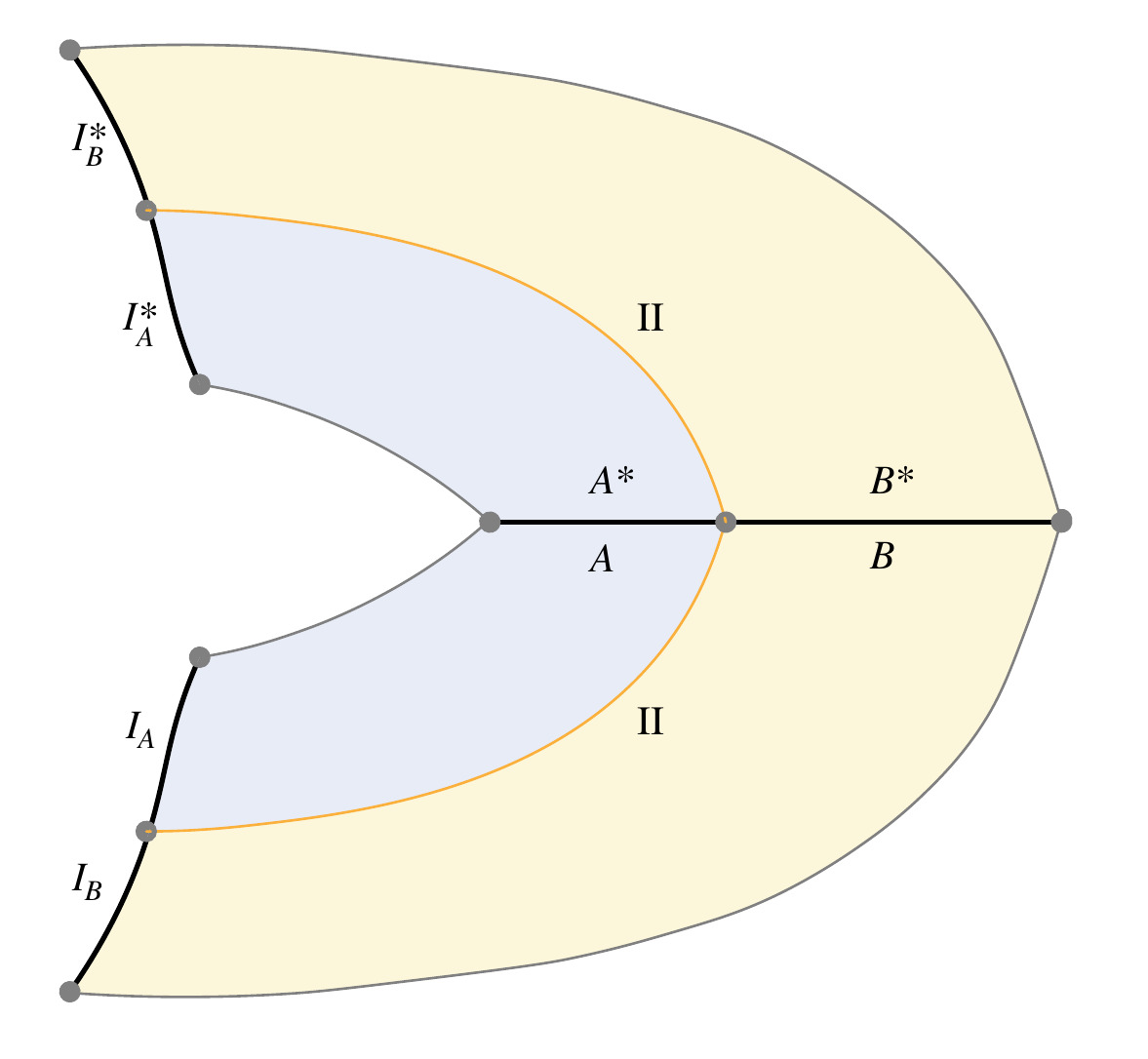}
\caption{Phase-II: 
The sandwich interval $C$ vanishes and both $A$ and $B$ are large to admit islands. 
Right: Purification of the interval $A \cup I_{R}(A)\cup B \cup I_{R}(B) $. 
}
\label{Fig:2}
\end{figure}

\begin{figure}[htp]
\centering
\includegraphics[width=0.8\textwidth]{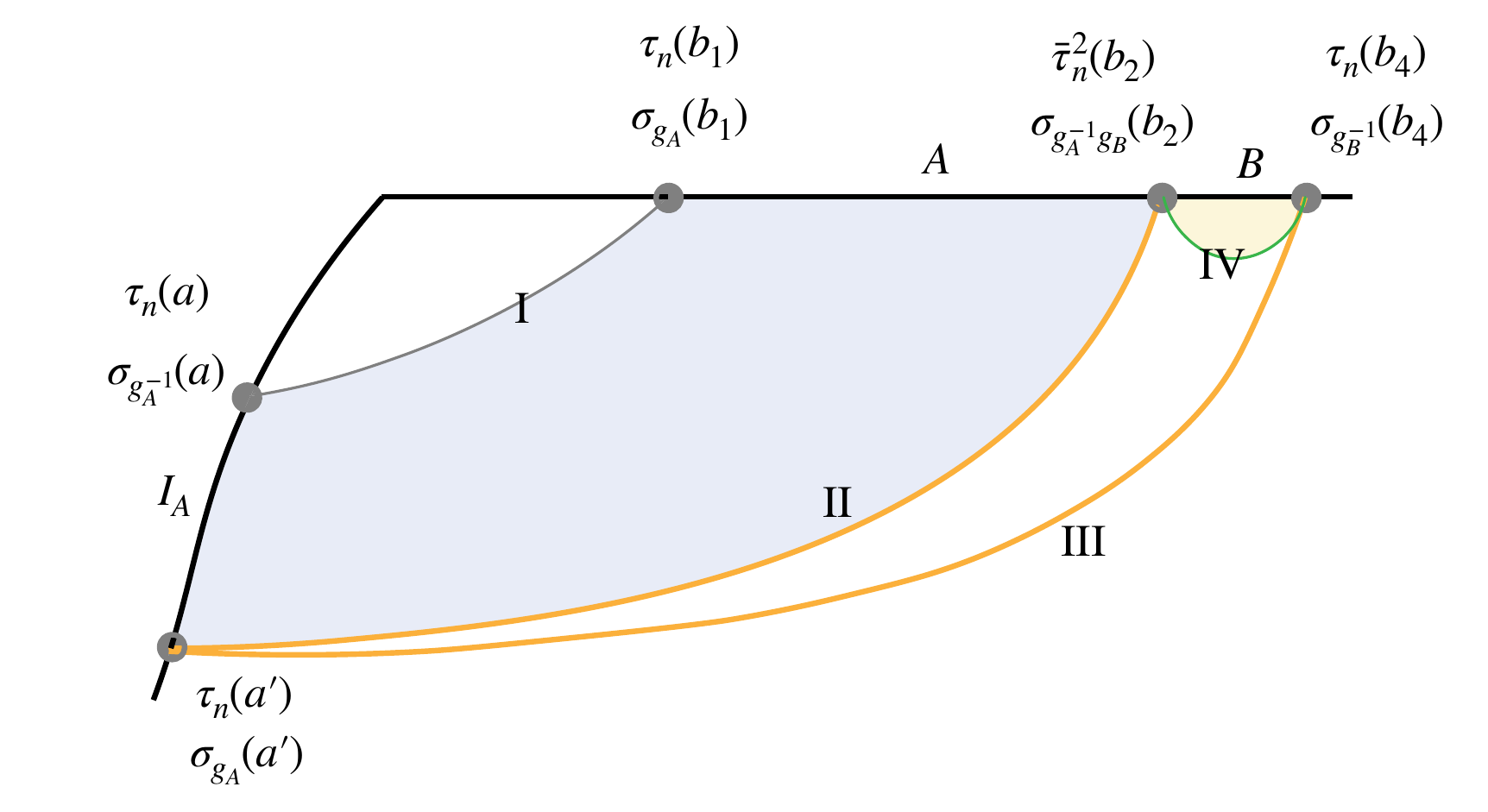}
\caption{Phase-III: B is too small to admit an island.}
\label{Fig:21}
\end{figure}

Our basic idea to extract the holographic scheme is that we first evaluate the multi-point correlation function of twist operators in large $c$ limit in vacuum coordinate $(x^+,x^-)$ and then express effective Renyi reflected entropy in terms of the combinations of modified minimal surfaces in $\text{AdS}_3$ bulk. 
Without loss of generality, we will perform the replica-trick computation \cite{Dutta:2019gen} of the effective Renyi  reflected entropy in large $c$ limit for phase-I (Fig.\ref{Fig:1}), phase-II (Fig.\ref{Fig:2}) and phase-III (Fig.\ref{Fig:21}) and extract the holographic scheme. 
Note that two disjoint intervals $A$ and $B$ of all phases we consider here are in proximity. 

To compute the Renyi reflected entropy, one should evaluate the effective Renyi reflected entropy \cite{Dutta:2019gen}
\begin{equation}\label{def_Snm}
    S_{R}^{(n,m)\rm eff}(A:B)=\frac{1}{1-n}\mathrm{Tr}\left(
    \rho_{AA^*}^{(m)}
    \right),
\end{equation}
where 
\begin{equation}
    \rho _{AA^*}^{(m)}\equiv \frac{1}{\mathrm{Tr}\rho_{AB}^{m}}\mathrm{Tr}_{BB^*}\rho_{AB}^{m/2}.
\end{equation}
In the limit $m\to 1$, eq. \eqref{def_Snm} is just Renyi reflected entropy $S_R^{(n)}$.

For phase-I where the sandwich interval $C$ is small and does not admit an island, the effective Renyi reflected entropy can be written in terms of the correlation functions of twist operators
\begin{equation}\label{rre}
\begin{split}
&S_{R}^{(n, m) \mathrm{eff}}(A \cup {I}_{R}(A): B \cup {I}_{R}(B))\\
=&\frac{1}{1-n} \log \frac{\left\langle\sigma_{g_{A}}\left(b_{1}\right) \sigma_{g_{A}^{-1}}(a) \sigma_{g_{A}^{-1}}\left(b_{2}\right) \sigma_{g_{B}}\left(b_{3}\right) \sigma_{g_{A} g_{B}^{-1}}\left(a^{\prime}\right)\sigma_{g_{B}^{-1}}\left(b_{4}\right) \sigma_{g_{B}}(a'') \right\rangle_{\text{CFT}^{\otimes mn}}}{\left\langle\sigma_{g_{m}}\left(b_{1}\right) \sigma_{g_{m}^{-1}}(a) \sigma_{g_{m}^{-1}}\left(b_{2}\right) \sigma_{g_{m}}\left(b_{3}\right)\sigma_{g_{m}^{-1}}\left(b_{4}\right) \sigma_{g_{m}}\left(a''\right)\right\rangle_{\text{CFT}^{\otimes m}}^{n}},
\end{split}
\end{equation}
where $I_{R}(A),I_{R}(B)$ are islands for reflected entropies of $A$ and $B$.
$\sigma_{g_A},\sigma_{g_A^{-1}},\sigma_{g_B},\sigma_{g_B^{-1}}$ are twist operators at the endpoints of the intervals. 
The twist operator $\sigma_{g_Bg_A^{-1}}$ gives the dominant contribution to the OPE \cite{Dutta:2019gen}
\begin{equation}
    \sigma_{g_A^{-1}}\sigma_{g_B}\to \sigma_{g_Bg_A^{-1}}+...
\end{equation}
The scaling dimensions of twist operators are
\begin{equation}
    h_{g_Ag_B^{-1}}=\frac{c}{12n}(n-1)(n+1)=2h_n,\
h_{g_A}=h_{g_A^{-1}}=\frac{cn(m^2-1)}{24m}=nh_m.
\end{equation}
$\sigma_{g_m}$ is $m$-twist operator on each $n$ replicas with scaling dimension
\begin{equation}
    h_m=\frac{c(m^2-1)}{24m}.
\end{equation}
In the large $c$ limit, the correlation function in eq.\eqref{rre} can be factorized as follow \cite{Hartman:2013mia}
\begin{equation}
\begin{split}
    &\left\langle\sigma_{g_{A}}\left(b_{1}\right) \sigma_{g_{A}^{-1}}(a) \sigma_{g_{A}^{-1}}\left(b_{2}\right) \sigma_{g_{B}}\left(b_{3}\right) \sigma_{g_{A} g_{B}^{-1}}\left(a^{\prime}\right)\sigma_{g_{B}^{-1}}\left(b_{4}\right) \sigma_{g_{B}}(a'')\right\rangle_{\text{CFT}^{\otimes m n}}\\
    \to&
    \left\langle\sigma_{g_{A}}\left(b_{1}\right) \sigma_{g_{A}^{-1}}(a)\right\rangle_{\text{CFT}^{\otimes m n}}
    \left\langle \sigma_{g_{A}^{-1}}\left(b_{2}\right) \sigma_{g_{B}}\left(b_{3}\right) \sigma_{g_{A} g_{B}^{-1}}\left(a^{\prime}\right)\right\rangle_{\text{CFT}^{\otimes m n}}
    \left\langle\sigma_{g_{B}^{-1}}\left(b_{4}\right) \sigma_{g_{B}}(a'')\right\rangle_{    \text{CFT}^{\otimes m n}},
    \end{split}
\end{equation}
\begin{equation}
\begin{split}
    &\left\langle\sigma_{g_{m}}\left(b_{1}\right) \sigma_{g_{m}^{-1}}(a) \sigma_{g_{m}^{-1}}\left(b_{2}\right) \sigma_{g_{m}}\left(b_{3}\right) \sigma_{g_{m}^{-1}}\left(b_{4}\right) \sigma_{g_{m}}(a'')\right\rangle_{\text{CFT}^{\otimes m}}\\
    \to&
    \left\langle\sigma_{g_{m}}\left(b_{1}\right) \sigma_{g_{m}^{-1}}(a)\right\rangle_{\text{CFT}^{\otimes m}}
    \left\langle \sigma_{g_{m}^{-1}}\left(b_{2}\right) \sigma_{g_{m}}\left(b_{3}\right) \right\rangle_{\text{CFT}^{\otimes m}}
    \left\langle\sigma_{g_{m}^{-1}}\left(b_{4}\right) \sigma_{g_{m}}(a'')\right\rangle_{\text{CFT}^{\otimes m}}.
    \end{split}
\end{equation}
Thus  the bulk Renyi reflected entropy becomes
\begin{equation}\label{rre1}
\begin{split}
&S_{R}^{(n, m) \mathrm{eff}}(A \cup {I}_{R}(A): B \cup {I}_{R}(B))\\
=& \frac{1}{1-n} \log \frac{\Omega_{a'}^{2h_{g_Ag_B^{-1}}}
d_{ab_1}^{-4h_{g_A}}
d_{b_2b_3}^{-4h_{g_A}+2h_{g_Ag_B^{-1}}}
d_{b_3a'}^{-2h_{g_Ag_B^{-1}}}
d_{b_2a'}^{-2h_{g_Ag_B^{-1}}}
d_{a''b_4}^{-4h_{g_B}}
}{\left(d_{ab_1}^{-4h_{m}}d_{b_3b_2}^{-4h_{m}}d_{a''b_4}^{-4h_{m}}\right)^n}+\text{const},
\end{split}
\end{equation}
where  
\begin{equation}
    d_{ab}=\sqrt{(x^+_a-x^+_b)(x^-_a-x^-_b)}
\end{equation}
is the distance in flat spacetime $\mathrm{d}s^2=-\mathrm{d}x^+\mathrm{d}x^-$.
The constant term is related to structure constants, which is subleading in the large $c$ limit \cite{Jain:2017aqk}.
Henceforth, we will drop this constant term.
Since $h_{g_A}=nh_m$, the terms in  the effective Renyi  reflected entropy \eqref{rre1} involving $h_{g_A},h_m$ cancel out. 
For phase-I, the only contribution is from the 3-point function
$\langle \sigma_{g_{A}^{-1}}(b_{2}) \sigma_{g_{B}}(b_{3}) \sigma_{g_{A} g_{B}^{-1}}(a^{\prime})\rangle$, and the effective Renyi reflected entropy is reduced to 
\begin{equation}\label{rre2}
\begin{split}
S_{R}^{(n, 1) \mathrm{eff}}(A \cup \operatorname{I}_{R}(A): B \cup \operatorname{I}_{R}(B))=&  \frac{1}{n-1}\log\left(  \frac{d_{b_3a'}d_{b_2a'}}{\Omega_{a'}d_{b_2b_3}}\right)^{2h_{g_Ag_B^{-1}}},
\end{split}
\end{equation}
which can be rewritten as
\begin{equation}\label{rre3}
\begin{split}
S_{R}^{(n, 1) \mathrm{eff}}(A \cup \operatorname{I}_{R}(A): B \cup \operatorname{I}_{R}(B))=&  \frac{c}{12}\left(1+\frac{1}{n}\right)\log\left(
\frac{d_{b_3a'}^2}{\Omega_{a'}\Omega_{b_3}}
\frac{d_{b_2a'}^2}{\Omega_{a'}\Omega_{b_2}}
\frac{\Omega_{b_2}\Omega_{b_3}}{d_{b_2b_3}^2}\right).
\end{split}
\end{equation}
Note that in $\mathrm{AdS}_3$ bulk, the term 
\begin{equation}
    \frac{c}{12}\left(1+\frac{1}{n}\right)\log\left(
\frac{d_{ab}^2}{\Omega_{a}\Omega_{b}}\right)\equiv \frac{\mathcal{A}(C^{(n)}_{ab})}{4G_N^{(3)}}
\end{equation}
is just the area of backreacting cosmic brane anchored at $a$ and $b$ \cite{Dong:2016fnf},  which is modified by conformal factor in the boundary 2D brane. 
The Renyi index $n$ is related to the tension of cosmic brane by
\begin{equation}
    T_n=\frac{n-1}{4nG_N^{(3)}}.
\end{equation}
Thus the effective Renyi reflected entropy \eqref{rre3} can be expressed as the combination of modified backreacting cosmic branes 
{\footnote{Note that for general black holes spacetime, our holographic scheme does not refer to a strict holographic construction but a geometric scheme for calculations. }}
up to a constant term
\begin{equation}\label{rre4}
\begin{split}
S_{R}^{(n, 1) \mathrm{eff}}(A \cup I_{R}(A): B \cup I_{R}(B))=&\frac{\mathcal{A}(C^{(n)}_{b_2a'})}{4G_N^{(3)}}+\frac{\mathcal{A}(C^{(n)}_{b_3a'})}{4G_N^{(3)}}-\frac{\mathcal{A}(C^{(n)}_{b_2b_3})}{4G_N^{(3)}}.
\end{split}
\end{equation}

Now let us consider the phase-II (Fig.\ref{Fig:2}) where the sandwich interval $C$ vanishes and the two twist operators $\sigma_{g^{-1}_A},\sigma_{g_B}$ merge into one $\sigma_{g^{-1}_Ag_B}$. 
Then the effective Renyi reflected entropy in large-$c$ limit is
\begin{equation}\label{re}
\begin{split}
&S_{R}^{(n, m) \mathrm{eff}}(A \cup {I}_{R}(A): B \cup {I}_{R}(B))\\
=&\frac{1}{1-n} \log \frac{\langle\sigma_{g_{A}}(b_{1})\sigma_{g_{A}^{-1}}(a)\rangle_{\text{CFT}^{\otimes mn}} \langle\sigma_{g_{A}^{-1}g_B}\left(b_{2}\right)\sigma_{g_{A}g_{B}^{-1}}\left(a^{\prime}\right)\rangle_{\text{CFT}^{\otimes mn}}\langle\sigma_{g_{B}^{-1}}(b_4)\sigma_{g_{B}}(a'')\rangle_{\text{CFT}^{\otimes mn}}}{\left\langle\sigma_{g_{m}}\left(b_{1}\right) \sigma_{g_{m}^{-1}}(a)\right\rangle_{\text{CFT}^{\otimes m}}^{n}\left\langle \sigma_{g_{m}^{-1}}\left(b_4\right) \sigma_{g_{m}}\left(a''\right)\right\rangle_{\text{CFT}^{\otimes m}}^{n}}\\
=&\frac{1}{1-n} \log\frac{(\Omega_{b_2}\Omega_{a'})^{2h_{g_Ag_b^{-1}}}d_{b_1a}^{-4h_{g_A}}d_{b_2a'}^{-4h_{g_Ag_B^{-1}}}d_{b_4a''}^{-4h_{g_B}}}{\left(d_{b_1a}^{-4h_m}d_{b_4a''}^{-4h_m}\right)^n}.
\end{split}
\end{equation}
Note that the terms involving scaling dimensions $h_{g_A},h_{g_B},h_m$ cancel out and thus the effective Renyi reflected entropy \eqref{re} becomes
\begin{equation}\label{re1}
\begin{split}
S_{R}^{(n, 1) \mathrm{eff}}(A \cup {I}_{R}(A): B \cup {I}_{R}(B))
=&\frac{1}{1-n} \log[(\Omega_{b_2}\Omega_{a'})d_{b_2a'}^{-2}]^{2h_{g_Ag_B^{-1}}}\\
=&\frac{c}{12}\left(1+\frac{1}{n}\right)\log\left(\frac{d_{b_2a'}^2}{\Omega_{b_2}\Omega_{a'}}\frac{d_{b_2a'}^2}{\Omega_{b_2}\Omega_{a'}}\right)\\
&=2\times\frac{\mathcal{A}(C^{(n)}_{b_2a'})}{4G_N^{(3)}},
\end{split}
\end{equation}
which is just twice the area of backreacting cosmic brane on EWCS between $A \cup {I}_{R}(A)$ and $B \cup {I}_{R}(B)$. 
In Fig.\ref{Fig:2}, we trace out the complement of the interval $A \cup I_{R}(A)\cup B \cup I_{R}(B) $ and then canonically purify the system by introducing another copy
$$
A^* \cup I_{R}(A^*)\cup B^* \cup I_{R}(B^*).
$$
It is clear that $2\mathcal{A}(C^{(n)}_{b_2a'})/4G_N^{(3)}$ is just the $n$-th Renyi reflected entropy between $A \cup {I}_{R}(A)$ and $B \cup {I}_{R}(B)$. 
Note that in the limit of the sandwich interval $C\to0$, the holographic scheme for phase-I \eqref{rre4} reduces to \eqref{re1} for phase-II. 

Now we can extract the contribution to effective Renyi reflected entropy from 2-point correlations.
The modified cosmic branes connecting two twist operators $\sigma_{g},\sigma_{g^{-1}}$ between island and radiation do not contribute to effective Renyi reflected entropy, while the cosmic branes connecting $\sigma_{g_Ag^{-1}_B}(b_1), \sigma_{g_Bg^{-1}_A}(a)$ contribute to $S_{R}^{(n,1)\rm eff}$ by
\begin{equation}
    2\times\frac{\mathcal{A}(C^{(n)}_{b_1a})}{4G_N^{(3)}}.
\end{equation}

Above discussion assumes that both intervals $A$ and $B$ are large enough to admit islands. 
Now let us consider the cases where the interval B is small and does not admit an island (Fig. \ref{Fig:21}). 
In this case, the twist operators $\sigma_{g_Ag_B^{-1}}(a'),\sigma_{g_B}(a'')$ in the island side merge into one twist operator $\sigma_{g_A}(a')$ and  the effective Renyi reflected entropy in large-$c$ limit and $m\to1$ becomes
\begin{equation}\label{rrre}
\begin{split}
&S_{R}^{(n, 1) \mathrm{eff}}(A \cup {I}_{R}(A): B \cup {I}_{R}(B))\\
=&\lim_{m\to1}\frac{1}{1-n} \log \frac{\langle\sigma_{g_{A}}(b_{1})\sigma_{g_{A}^{-1}}(a)\rangle_{\text{CFT}^{\otimes mn}} \langle\sigma_{g_{A}^{-1}g_B}\left(b_{2}\right)\sigma_{g_{A}}\left(a^{\prime}\right)\sigma_{g_{B}^{-1}}(b_4)\rangle_{\text{CFT}^{\otimes m n}}}{\left\langle\sigma_{g_{m}}\left(b_{1}\right) \sigma_{g_{m}^{-1}}(a)\right\rangle_{\text{CFT}^{\otimes m}}^{n}\left\langle \sigma_{g_{m}^{-1}}\left(b_4\right) \sigma_{g_{m}}\left(a'\right)\right\rangle_{\text{CFT}^{\otimes m}}^{n}}\\
=&\lim_{m\to1}\frac{1}{1-n} \log\frac{(\Omega_{b_2})^{2h_{g_Ag_B^{-1}}}d_{b_1a}^{-4h_{g_A}}d_{b_2a'}^{-2h_{g_Ag_B^{-1}}}d_{b_4a'}^{-4h_{g_B}+2h_{g_Ag_B^{-1}}}d_{b_2b_4}^{-2h_{g_Ag_B^{-1}}}}{\left(d_{b_1a}^{-4h_m}d_{b_4a'}^{-4h_m}\right)^n}\\
=&\frac{c}{12}\left(1+\frac{1}{n}\right)\log [d_{b_2a'}^{2}d_{b_4a'}^{-2}d_{b_2b_4}^{2}\Omega_{b_2}^{-2}]\\
=&\frac{c}{12}\left(1+\frac{1}{n}\right)\log \left(
\frac{d_{b_2a'}^2}{\Omega_{b_2}\Omega_{a'}}\frac{d_{b_2b_4}^2}{\Omega_{b_2}\Omega_{b_4}}\frac{\Omega_{a'}\Omega_{b_4}}{d_{b_4a'}^2}\right)\\
=&\frac{\mathcal{A}(C^{(n)}_{b_2a'})}{4G_N^{(3)}}+\frac{\mathcal{A}(C^{(n)}_{b_2b_4})}{4G_N^{(3)}}-\frac{\mathcal{A}(C^{(n)}_{b_4a'})}{4G_N^{(3)}}.
\end{split}
\end{equation}
From eqs. \eqref{rre4} and \eqref{rrre}, we can extract the contribution to effective Renyi reflected entropy  from 3-point connection. 
That is, for 3-point connection with twist operators $\sigma_{g_B}(b_1)$, $\sigma_{g^{-1}_A}(b_2)$ and $\sigma_{g_Ag^{-1}_B}(b_3)$, the contribution to effective Renyi reflected entropy is
$$
\frac{\mathcal{A}(C^{(n)}_{b_1b_3})}{4G_N^{(3)}}+\frac{\mathcal{A}(C^{(n)}_{b_2b_3})}{4G_N^{(3)}}-\frac{\mathcal{A}(C^{(n)}_{b_1b_2})}{4G_N^{(3)}}.
$$

\subsection{Holographic calculation scheme for effective Renyi reflected entropy }
Now we conclude our holographic scheme for effective Renyi reflected entropy in large $c$ limit. 

\begin{enumerate}
    \item  Connect the boundaries of radiation and the corresponding island by modified backreacting cosmic brane in $\mathrm{AdS}_3$ bulk. 
    \item 
    The 2-point connection of twist operators $\sigma_{g},\sigma_{g^{-1}}$ between island and radiation does not contribute, while the contribution from the 2-point connection of twist operators $\sigma_{g_Ag^{-1}_B}(b), \sigma_{g_Bg^{-1}_A}(a)$ is given by twice the area of the backreacting cosmic brane in the unit of $4G_N^{(3)}$
\begin{equation}
    2\times\frac{\mathcal{A}(C^{(n)}_{ba})}{4G_N^{(3)}}\equiv \mathcal{A}(C^{(n)})_{\text{2-point}}.
\end{equation}
\item The contribution from 3-point connection of twist operators $\sigma_{g_B}(b_1),\sigma_{g^{-1}_A}(b_2),\sigma_{g_Ag^{-1}_B}(b_3)$ is given by a combination of the areas of modified backreacting cosmic branes up to a constant term
\begin{equation}
    \frac{\mathcal{A}(C^{(n)}_{b_1b_3})}{4G_N^{(3)}}+\frac{\mathcal{A}(C^{(n)}_{b_2b_3})}{4G_N^{(3)}}-\frac{\mathcal{A}(C^{(n)}_{b_1b_2})}{4G_N^{(3)}}\equiv \mathcal{A}(C^{(n)})_{\text{3-point}}.
\end{equation}
\item 
The effective Renyi reflected entropy in large-$c$ limit then is given by the sum of 2-point and 3-point contributions
\begin{equation}\label{SR_eff}
    S^{(n,1)\text{eff}}_R= \mathcal{A}(C^{(n)})_{\text{2-point}}+\mathcal{A}(C^{(n)})_{\text{3-point}}.
\end{equation}
\end{enumerate}

\section{Effective entanglement negativity}\label{sec3}
In this section, we will develop a holographic scheme to calculate effective entanglement negativity, which is similar to that for effective Renyi reflected entropy proposed in Sec.\ref{sec2}.
And we will see that the effective entanglement negativity can also be expressed in terms of the combination of the areas of modified backreacting cosmic branes in $\mathrm{AdS}_3$ bulk. 

For phase-I where the sandwich interval $C$ is too small to admit an island,
the effective Renyi entanglement negativity can be written in terms of the correlation functions of twist operators \cite{Calabrese:2012ew,Calabrese:2012nk}. 
In the large $c$ limit, the effective Renyi entanglement negativity becomes
\begin{equation}
\begin{split}
&\mathcal{E}^{(n_e)\mathrm{eff}}\left(A \cup I_{\mathcal{E}}(A): B \cup I_{\mathcal{E}}(B)\right)\\
=& \log \left\langle\tau_{n_{e}}\left(b_{1}\right) \bar{\tau}_{n_{e}}(a) \bar{\tau}_{n_{e}}\left(b_{2}\right) \bar{\tau}_{n_{e}}\left(b_{3}\right) \tau_{n_{e}}^{2}\left(a^{\prime}\right)\right\rangle\\
=&\log \left\langle\tau_{n_{e}}\left(b_{1}\right) \bar{\tau}_{n_{e}}(a)\rangle\langle \bar{\tau}_{n_{e}}\left(b_{2}\right) \bar{\tau}_{n_{e}}\left(b_{3}\right) \tau_{n_{e}}^{2}\left(a^{\prime}\right)\right\rangle\\
=&\log \Omega_{a'}^{2h_{\tau^2_{n_e}}}
(\Omega_{b_1}\Omega_{a}\Omega_{b_2}\Omega_{b_3})^{h_{\tau_{n_e}}}
d_{ab_1}^{-4h_{\tau_{n_e}}}d_{b_3a'}^{-2h_{\tau^2_{n_e}}}d_{b_2a'}^{-2h_{\tau^2_{n_e}}}d_{b_3b_2}^{-4h_{\tau_{n_e}}+2h_{\tau^2_{n_e}}},
\end{split}
\end{equation}
where $I_\mathcal{E}$ denotes the island for entanglement negativity, and the scaling dimensions are given by \cite{Calabrese:2012nk}
\begin{equation}
h_{\tau_{n_{e}}}=\frac{c}{24}\left(n_{e}-\frac{1}{n_{e}}\right), \quad h_{\tau_{n_{e}}^{2}}=\frac{c}{12}\left(\frac{n_{e}}{2}-\frac{2}{n_{e}}\right).
\end{equation}
Taking the Renyi index $n_e\to1$, the effective entanglement negativity is 
\begin{equation}
\begin{split}
&\mathcal{E}^{\mathrm{eff}}\left(A \cup I_{\mathcal{E}}(A): B \cup I_{\mathcal{E}}(B)\right)\\
=&\lim_{n_e\to1}\log \left[\Omega_{a'}^{2h_{\tau^2_{n_e}}}
(\Omega_{b_1}\Omega_{a}\Omega_{b_2}\Omega_{b_3})^{h_{\tau_{n_e}}}
d_{ab_1}^{-4h_{\tau_{n_e}}}d_{b_3a'}^{-2h_{\tau^2_{n_e}}}d_{b_2a'}^{-2h_{\tau^2_{n_e}}}d_{b_3b_2}^{-4h_{\tau_{n_e}}+2h_{\tau^2_{n_e}}}\right]\\
=&\frac{c}{8}\log \left[\Omega_{a'}^{-2}
d_{b_3a'}^{2}d_{b_2a'}^{2}d_{b_3b_2}^{-2}\right]\\
=&\frac{c}{8}\log \left[\frac{d_{b_3a'}^{2}}{\Omega_{a'}\Omega_{b_3}}
\frac{d_{b_2a'}^{2}}{\Omega_{a'}\Omega_{b_2}}\frac{\Omega_{b_2}\Omega_{b_3}}{d_{b_2b_3}^{2}}
\right].
\end{split}
\end{equation}
Note that in AdS$_3$, the term 
\begin{equation}
\frac{c}{8}\log \left[\frac{d_{ab}^{2}}{\Omega_{a}\Omega_{b}}\right]\equiv\frac{1}{2}\frac{\mathcal{A}(C^{(n=1/2)}_{ab})}{4G_N^{(3)}}    
\end{equation}
is half the area of the backreacting cosmic brane of order 1/2 anchored at $a$ and $b$, which is modified by conformal factor in the boundary 2D brane.  
Thus the effective entanglement negativity in large $c$ limit can be  expressed as the combination of modified backreacting cosmic branes up to a constant term 
\begin{equation}
\begin{split}
&\mathcal{E}^{\mathrm{eff}}\left(A \cup I_{\mathcal{E}}(A): B \cup I_{\mathcal{E}}(B)\right)
=\frac{1}{2}\left(\frac{\mathcal{A}(C^{(1/2)}_{b_2a'})}{4G_N^{(3)}}+\frac{\mathcal{A}(C^{(1/2)}_{b_3a'})}{4G_N^{(3)}}-\frac{\mathcal{A}(C^{(1/2)}_{b_2b_3})}{4G_N^{(3)}}\right).
\end{split}
\end{equation}
For phase-II where the sandwich interval $C$ vanishes and two twist operators $\bar{\tau}_n(b_2),\bar{\tau}_n(b_3)$ in the radiation side merge into one $\bar{\tau}^2_n(b_2)$, the effective entanglement negativity in large $c$ limit becomes
\begin{equation}
\begin{split}
&\mathcal{E}^{\mathrm{eff}}\left(A \cup I_{\mathcal{E}}(A): B \cup I_{\mathcal{E}}(B)\right)\\
=& \lim_{n_e\to1}\log \left\langle\tau_{n_{e}}\left(b_{1}\right) \bar{\tau}_{n_{e}}(a) \bar{\tau}_{n_{e}}^2\left(b_{2}\right)  \tau_{n_{e}}^{2}\left(a^{\prime}\right)
\tau_{n_{e}}\left(b_{4}\right) \bar{\tau}_{n_{e}}(a'')
\right\rangle\\
=& \frac{c}{8}\log \left(\Omega_{b_2}^{-2}\Omega_{a'}^{-2}d_{b_2a'}^{4}\right)\\
=&\frac{1}{2}\times 2\frac{\mathcal{A}(C^{(1/2)}_{b_2a'})}{4G_N^{(3)}}.
\end{split}
\end{equation}
For phase-III where the interval $B$ is too small to admit an island and two twist operators $\tau^2_n(a'),\bar{\tau}_n(a'')$ in the island side merge into one $\tau_n(a')$,
the effective entanglement negativity in large $c$ limit is
\begin{equation}
\begin{split}
&\mathcal{E}^{\mathrm{eff}}\left(A \cup I_{\mathcal{E}}(A): B \cup I_{\mathcal{E}}(B)\right)\\
=& \lim_{n_e\to1}\log \left\langle\tau_{n_{e}}\left(b_{1}\right) \bar{\tau}_{n_{e}}(a) \bar{\tau}_{n_{e}}^2\left(b_{2}\right)  \tau_{n_{e}}\left(a^{\prime}\right)
\tau_{n_{e}}\left(b_{4}\right)
\right\rangle\\
=& \lim_{n_e\to1}\log \left\langle\bar{\tau}_{n_{e}}^2\left(b_{2}\right)  \tau_{n_{e}}\left(a^{\prime}\right)
\tau_{n_{e}}\left(b_{4}\right)
\right\rangle\\
=& \frac{c}{8}\log \left[\Omega_{b_2}^{-2}
d_{b_2a'}^{2}d_{b_2b_4}^{2}d_{b_4a'}^{-2}\right]\\
=&\frac{1}{2}\left(\frac{\mathcal{A}(C^{(1/2)}_{b_2a'})}{4G_N^{(3)}}+\frac{\mathcal{A}(C^{(1/2)}_{b_2b_4})}{4G_N^{(3)}}-\frac{\mathcal{A}(C^{(1/2)}_{b_4a'})}{4G_N^{(3)}}\right).
\end{split}
\end{equation}
\subsection{Holographic calculation scheme for effective entanglement negativity }
Now we conclude our holographic scheme for effective entanglement negativity in large $c$ limit. 

\begin{enumerate}
    \item 
     Connect the endpoints of the island and the radiation by modified backreacting cosmic brane of order 1/2 in $\mathrm{AdS}_3$ bulk.
     \item
     The 2-point connections of twist operators $\tau_{n_e},\bar{\tau}_{n_e}$ between island and radiation has no contribution.
     For 2-point connection of twist operators $\tau^2_{n_e},\bar{\tau}^2_{n_e}$,
write down the combination of modified backreacting cosmic branes of $1/2$ order 
$$
2\times\frac{\mathcal{A}(C^{(1/2)}_{b_1a})}{4G_N^{(3)}}\equiv \mathcal{A}(C^{(1/2)})_{\text{2-point}}
$$
\item
For 3-point connection of twist operators $\tau_{n_e}(b_1),\tau_{n_e}(b_2),\bar{\tau}^2_{n_e}(b_3)$, 
write down the combination of modified backreacting cosmic branes of $1/2$ order up to a constant term 
$$
\frac{\mathcal{A}(C^{(1/2)}_{b_1b_3})}{4G_N^{(3)}}+\frac{\mathcal{A}(C^{(1/2)}_{b_2b_3})}{4G_N^{(3)}}-\frac{\mathcal{A}(C^{(1/2)}_{b_1b_2})}{4G_N^{(3)}}\equiv \mathcal{A}(C^{(1/2)})_{\text{3-point}}.
$$
\item
     The effective entanglement negativity in large-$c$ limit then is given by half the sum of 2-point and 3-point contributions
\begin{equation}\label{HS_negativity}
    \mathcal{E}^{\text{eff}}=\frac{1}{2}\left(\mathcal{A}(C^{(1/2)})_{\text{2-point}}+\mathcal{A}(C^{(1/2)})_{\text{3-point}}\right).
\end{equation}
\end{enumerate}

\section{Consistency check of island formula of entanglement negativity  for general 2D eternal black holes}\label{sec4}
In this section, we will apply our holographic schemes for effective Renyi reflected entropy and effective entanglement negativity to study the consistency condition for entanglement negativity. 

The quantum corrected entanglement negativity can be obtained by extremizing the generalized entanglement negativity \cite{Kudler-Flam:2018qjo}. 
Coupled to semiclassical gravity, the generalized entanglement negativity should be modified by including the island \cite{KumarBasak:2020ams}
\begin{equation}\label{p3}
\begin{split}
    \mathcal{E}(A: B)=&
    \min(\mathrm{ext}_{Q''}\{\mathcal{E}_{\rm gen}(A:B)\}),
\end{split}
\end{equation}
where 
\begin{equation}
    \mathcal E_{\rm gen}(A:B)= 
    \frac{\mathcal{A}^{(1 / 2)}\left(Q^{\prime \prime}=\partial {I}_{ \mathcal{E}}(A) \cap \partial {I}_{ \mathcal{E}}(B)\right)}{4 G_{N}}+\mathcal{E}^{\text {eff }}\left(A \cup I _{\mathcal{E}}(A): B \cup I_{\mathcal{E}}(B)\right),
\end{equation}
where $I_{\mathcal E}$ denotes the possible island for entanglement negativity, which obeys $I_\mathcal{E}(A)\cup I_\mathcal{E}(B)=I(A\cup B)$ with $I$ being the island for entanglement entropy. 
$\mathcal A^{(1/2)}$ is the area of backreacting cosmic brane. 
The index 1/2 is related to the tension of cosmic brane. 
$Q''$ that minimizes $\mathcal E_{\rm gen}(A:B)$ is the quantum entanglement wedge cross section (QEWCS) for entanglement negativity. 

An alternative  proposal to calculate the entanglement negativity is to extremize the generalized Renyi reflected entropy of order 1/2 \cite{KumarBasak:2020ams}
\begin{equation}\label{p1}
2\mathcal{E}(A: B)=\text{min}(\text{ext}_{Q'}\{S_{R \text { gen }}^{(1 / 2)}(A: B)\}),
\end{equation}
where the generalized Renyi reflected entropy of order 1/2 is 
\begin{equation}
S_{R \text { gen }}^{(1 / 2)}(A: B)=\frac{2\mathcal{A}^{(1 / 2)}\left(Q'= \partial {I}_{R}(A) \cap \partial {I}_{R}(B)\right)}{4 G_{N}}+S_{R \text { eff }}^{(1 / 2)}\left(A \cup {I}_{R}(A): B \cup {I}_{R}(B)\right),
\end{equation}
$Q'$ is QEWCS for reflected entropy. 
$I_R$ denotes the island for reflected entropy.

On the other hand, when the interval between $A$ and $B$ is in proximity, the entanglement negativity can be obtained by extremzing the combination of the generalized Renyi entropy of order 1/2 \cite{KumarBasak:2020ams}
\begin{equation}\label{p2}
\mathcal{E}_{\mathrm{gen}}(A: B)=\frac{1}{2}\left[S_{\rm g e n}^{(1 / 2)}(A \cup C)+S_{\rm g e n}^{(1 / 2)}(B \cup C)-S_{\rm g e n}^{(1 / 2)}(A \cup B \cup C)-S_{\rm g e n}^{(1 / 2)}(C)\right],
\end{equation}
where the generalized Renyi entropy of order 1/2 is 
\begin{equation}
    S_{\rm  g e n}^{(1 / 2)}(A)=\frac{\mathcal{A}^{(1/2)}(\partial I (A))}{4G_N}+S^{(1/2)}_{\text{eff}}[A\cup I(A)],
\end{equation}
$S^{(1/2)}_{\text{eff}}[A\cup I(A)]$ is the effective Renyi entropy of order 1/2. 
Proposals in eqs.\eqref{p3},\eqref{p1} and \eqref{p2} provide a consistency check for island formulae of entanglement negativity.
In Ref.\cite{KumarBasak:2020ams}, the author checked the consistency condition for eternal JT black holes coupled to CFT matters. 
In this section, we will take a step further and study the consistency condition for general 2D black holes coupled to CFT matters by utilizing our holographic calculation scheme for effective Renyi reflected entropy and effective entanglement negativity. 
We will see the exact match of the combinations of modified cosmic branes in $\text{AdS}_3$ bulk from proposals in eqs.\eqref{p3},\eqref{p1} and \eqref{p2}.
Our results then indicate that the island for reflected entropy and entanglement negativity match with each other for general 2D black holes coupled to CFT matters.

\begin{figure}[htp]
\centering
\includegraphics[width=0.8\textwidth]{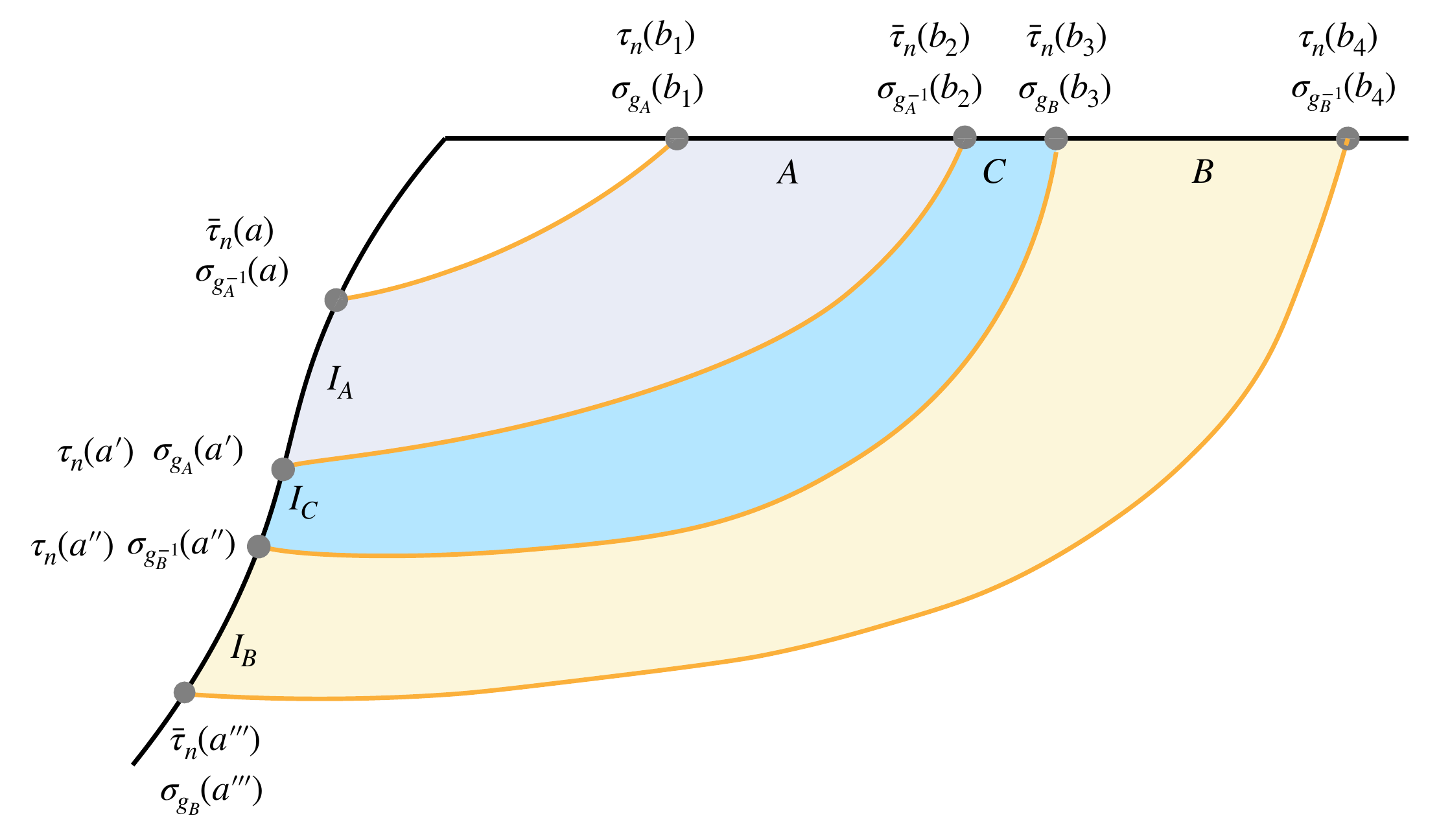}
\caption{Phase-IV: The sandwich interval $C$ is large enough to admit an island.}
\label{Fig:3}
\end{figure}

In order to prove the consistency condition, we will first develop a holographic scheme for the calculations of effective Renyi entropy of order $n$.  
In the case that the interval $[x_1,x_2]$ is large and admits an island, the effective Renyi entroy is
\begin{equation}
S^{(n)}_{\text{eff}}([x_1,x_2]\cup [\partial I_{x_1},\partial I_{x_2}])=\frac{1}{1-n} \log \left\langle\sigma\left(x_{1}\right)
\tilde{\sigma}\left(\partial I_{x_{1}} \right)
\tilde{\sigma}\left(x_{2} \right)
\sigma\left(\partial I_{x_{2}}\right)
\right\rangle_{\mathcal{C}^{n} / \mathbb{Z}_{n}},
\end{equation}
where $\sigma,\tilde{\sigma}$ are twist operators with scaling dimension
\begin{equation}
d_{n}=\frac{c}{12} \frac{(n-1)(n+1)}{n}.
\end{equation}
When the interval $[x_1,x_2]$ is large enough, the 4-point function can be factorized into two 2-point functions,
\begin{equation}
\left\langle\sigma\left(x_{1}\right)
\tilde{\sigma}\left(\partial I_{x_{1}} \right)
\tilde{\sigma}\left(x_{2} \right)
\sigma\left(\partial I_{x_{2}}\right)
\right\rangle\to\left\langle\sigma\left(x_{1}\right)
\tilde{\sigma}\left(\partial I_{x_{1}} \right)\rangle
\langle\tilde{\sigma}\left(x_{2} \right)
\sigma\left(\partial I_{x_{2}}\right)
\right\rangle,
\end{equation}
then the effective Renyi entropy becomes
\begin{equation}
\begin{split}
&S^{(n)}_{\text{eff}}([x_1,x_2]\cup [\partial I_{x_1},\partial I_{x_2}])\\
=& \frac{1}{1-n}\log \left((\Omega_{x_{1}}\Omega_{\partial I_{x_{1}}}\Omega_{x_{2}}\Omega_{\partial I_{x_{2}}})^{d_n}\left\langle\sigma\left(x_{1}\right)
\tilde{\sigma}\left(\partial I_{x_{1}} \right)\rangle
\langle\tilde{\sigma}\left(x_{2} \right)
\sigma\left(\partial I_{x_{2}}\right)
\right\rangle\right)\\
=&\frac{c}{12} \frac{n+1}{n}\log 
\frac{d(x_1,\partial I_{x_1})^{2}d(x_2,\partial I_{x_2})^{2}}{\Omega_{x_{1}}\Omega_{\partial I_{x_{1}}}\Omega_{x_{2}}\Omega_{\partial I_{x_{2}}}}\\
=&\frac{\mathcal{A}(C^{(n)}_{x_{1}\partial I_{x_{1}}})}{4G_N^{(3)}}+\frac{\mathcal{A}(C^{(n)}_{x_{2}\partial I_{x_{2}}})}{4G_N^{(3)}}.
\end{split}
\end{equation}
and we have expressed the effective Renyi entropy in terms of the combination of modified backreacting cosmic branes in $\mathrm{AdS}_3$ bulk.

In the case that the interval $[x_1,x_2]$ is small and admit no island, the  effective Renyi entropy is 
\begin{equation}
\begin{split}
S^{(n)}_{\text{eff}}([x_1,x_2]\cup [\partial I_{x_1},\partial I_{x_2}])=& S^{(n)}_{\text{eff}}([x_1,x_2])\\
=&\frac{c}{12} \frac{n+1}{n}\log 
\frac{d(x_1,x_2)^{2}}{\Omega(x_1)\Omega(x_2)}\\
=&\frac{\mathcal{A}(C^{(n)}_{x_1x_2})}{4G_N^{(3)}}.
\end{split}
\end{equation}
With the holographic scheme for effective Renyi (reflected) entropy and effective entanglement negativity in hand, now we give a proof of the consistency condition.
Without loss of generality, we will restrict to the four phases (see Figs.\ref{Fig:1}-\ref{Fig:3}). 

\subsection{Phase-I}
Consider phase-I where the sandwich interval $C$ is small and does not admit an island.
According to eq.\eqref{HS_negativity}, 
the generalized entanglement negativity from proposal \eqref{p3} is
\begin{equation}\label{Egen_p3}
\begin{split}
    \mathcal{E}_{\mathrm{gen}}(A: B)=&\frac{\mathcal{A}^{1/2}(a'_{\mathcal E})}{4G_N^{(2)}}+\mathcal{E}^{\text{eff}}\\
    =&\frac{\mathcal{A}^{1/2}(a'_{\mathcal E})}{4G_N^{(2)}}+\frac{1}{2}\left(\frac{\mathcal{A}(C^{(1/2)}_{\mathcal E,\text{IV}})}{4G_N^{(3)}}+\frac{\mathcal{A}(C^{(1/2)}_{\mathcal E,\text{II}})}{4G_N^{(3)}}-\frac{\mathcal{A}(C^{(1/2)}_{\mathcal E,\text{III}})}{4G_N^{(3)}}\right),
    \end{split}
\end{equation}
where the subscript $\mathcal E$ indicates that the quantity is given by island for entanglement negativity.

Now let us turn to proposal eq.\eqref{p1}.
According to eq.\eqref{SR_eff}, the effective Renyi reflected entropy is given by
\begin{equation}
\begin{split}
    S^{(1/2,1)\text{eff}}_R=& \mathcal{A}(C_{R}^{(1/2)})_{\text{2-point}}+\mathcal{A}(C^{(1/2)}_R)_{\text{3-point}}\\
    =&\mathcal{A}(C_R^{(1/2)})_{\text{3-point}}=\frac{\mathcal{A}(C^{(1/2)}_{R,\text{IV}})}{4G_N^{(3)}}+\frac{\mathcal{A}(C^{(1/2)}_{R,\text{II}})}{4G_N^{(3)}}-\frac{\mathcal{A}(C^{(1/2)}_{R,\text{III}})}{4G_N^{(3)}},
    \end{split}
\end{equation}
in which the subscript $R$ indicates that the cosmic branes are those evaluating the Renyi reflected entropy with island $I_R$ for reflected entropy.
And the generalized Renyi reflected  entropy is
\begin{equation}
    S_{R~\rm gen}^{(1/2)}=\frac{2\mathcal{A}^{1/2}(a'_{R})}{4G_N^{(2)}}+\frac{\mathcal{A}(C^{(1/2)}_{R,\text{IV}})}{4G_N^{(3)}}+\frac{\mathcal{A}(C^{(1/2)}_{R,\text{II}})}{4G_N^{(3)}}-\frac{\mathcal{A}(C^{(1/2)}_{R,\text{III}})}{4G_N^{(3)}}, 
\end{equation}
and the entanglement negativity $\mathcal {E}(A:B)$ is given by half the minimal $S^{(1/2)}_{R~\rm gen}$.

Thirdly, the generalized entanglement negativity can be also given by eq.\eqref{p2}.
Since we have
\begin{equation}
S_{\text{gen}}^{(1/2)}(A\cup C)=\frac{\mathcal{A}^{(1/2)}(a)}{4G_N^{(2)}}+\frac{\mathcal{A}^{(1/2)}(a')}{4G_N^{(2)}}+\frac{\mathcal{A}(C^{(1/2)}_{\text{I}})}{4G_N^{(3)}}+\frac{\mathcal{A}(C^{(1/2)}_{\text{IV}})}{4G_N^{(3)}},
\end{equation}
\begin{equation}
S_{\text{gen}}^{(1/2)}(B\cup C)=\frac{\mathcal{A}^{(1/2)}(a')}{4G_N^{(2)}}+\frac{\mathcal{A}^{(1/2)}(a'')}{4G_N^{(2)}}+\frac{\mathcal{A}(C^{(1/2)}_{\text{II}})}{4G_N^{(3)}}+\frac{\mathcal{A}(C^{(1/2)}_{\text{V}})}{4G_N^{(3)}},
\end{equation}
\begin{equation}
S_{\text{gen}}^{(1/2)}(A\cup B\cup C)=\frac{\mathcal{A}^{(1/2)}(a)}{4G_N^{(2)}}+\frac{\mathcal{A}^{(1/2)}(a'')}{4G_N^{(2)}}+\frac{\mathcal{A}(C^{(1/2)}_{\text{I}})}{4G_N^{(3)}}+\frac{\mathcal{A}(C^{(1/2)}_{\text{V}})}{4G_N^{(3)}},
\end{equation}
\begin{equation}
S_{\text{gen}}^{(1/2)}(C)=\frac{\mathcal{A}(C^{(1/2)}_{\text{III}})}{4G_N^{(3)}},
\end{equation}
the generalized entanglement negativity for phase-I is 
\begin{equation}\label{pp1}
\begin{split}
    \mathcal{E}_{\mathrm{gen}}(A: B)
    =&\frac{\mathcal{A}^{1/2}(a')}{4G_N^{(2)}}+\frac{1}{2}\left(\frac{\mathcal{A}(C^{(1/2)}_{\text{IV}})}{4G_N^{(3)}}+\frac{\mathcal{A}(C^{(1/2)}_{\text{II}})}{4G_N^{(3)}}-\frac{\mathcal{A}(C^{(1/2)}_{\text{III}})}{4G_N^{(3)}}\right),
    \end{split}
\end{equation}
where the location of $a'$ (and thus the cosmic branes $C^{(1/2)}$) is determined by the island for entanglement entropy.

We see that $\mathcal E_{\rm gen}(A:B)$ in eqs. \eqref{Egen_p3} and \eqref{pp1} share the same expressions with $S^{(1/2)}_{R~\rm gen}/2$ except for the location of $a'$. 
However, after minimization, the three expressions give exactly the same result $\mathcal E_{\rm gen}(A:B)$, as well as the same $a'$, which leads to $I_R=I_{\cal E}=I$.
Below, we will suppress the subscripts $R$ and $\mathcal E$, but keep in mind that the $a'$ (and thus $C^{(n)}$) has various origins.

\subsection{Phase-II}
For phase-II (Fig.\ref{Fig:2}) where the sandwich interval $C$ vanishes, the generalized entanglement negativity from proposal eq.\eqref{p3} is
\begin{equation}\label{Egen_II1}
\begin{split}
    \mathcal{E}_{\mathrm{gen}}(A: B)=&\frac{\mathcal{A}^{1/2}(a')}{4G_N^{(2)}}+\mathcal{E}^{\text{eff}}\\
    =&\frac{\mathcal{A}^{1/2}(a')}{4G_N^{(2)}}+\frac{1}{2}\mathcal{A}(C^{(1/2)})_{\text{2-point}}\\
    =&\frac{\mathcal{A}^{1/2}(a')}{4G_N^{(2)}}+\frac{\mathcal{A}(C^{(1/2)}_{\text{II}})}{4G_N^{(3)}}.
    \end{split}
\end{equation}

Secondly, the effective Renyi reflected entropy is
\begin{equation}
\begin{split}
    S^{(1/2,1)\text{eff}}_R=& \mathcal{A}(C^{(1/2)})_{\text{2-point}}+\mathcal{A}(C^{(1/2)})_{\text{3-point}}\\
    =&\mathcal{A}(C^{(1/2)})_{\text{2-point}}=\frac{2\mathcal{A}(C^{(1/2)}_{\text{II}})}{4G_N^{(3)}},
    \end{split}
\end{equation}
then the entanglement negativity is given by half the minimal generalized Renyi reflected entropy according to eq.\eqref{p1}
\begin{equation}\label{ppp2}
\begin{split}
    \frac12S^{(1/2)}_{R~\mathrm{gen}}(A: B)=&\frac{\mathcal{A}^{1/2}(a')}{4G_N^{(2)}}+\frac{1}{2}\left(\mathcal{A}(C^{(1/2)})_{\text{2-point}}+\mathcal{A}(C^{(1/2)})_{\text{3-point}}\right)\\
    =&\frac{\mathcal{A}^{1/2}(a')}{4G_N^{(2)}}+\frac{\mathcal{A}(C^{(1/2)}_{\text{II}})}{4G_N^{(3)}}.
    \end{split}
\end{equation}

Thirdly, since we have 
\begin{equation}
S_{\text{gen}}^{(1/2)}(A)=\frac{\mathcal{A}^{(1/2)}(a)}{4G_N^{(2)}}+\frac{\mathcal{A}^{(1/2)}(a')}{4G_N^{(2)}}+\frac{\mathcal{A}(C^{(1/2)}_{\text{I}})}{4G_N^{(3)}}+\frac{\mathcal{A}(C^{(1/2)}_{\text{II}})}{4G_N^{(3)}},
\end{equation}
\begin{equation}
S_{\text{gen}}^{(1/2)}(B)=\frac{\mathcal{A}^{(1/2)}(a')}{4G_N^{(2)}}+\frac{\mathcal{A}^{(1/2)}(a'')}{4G_N^{(2)}}+\frac{\mathcal{A}(C^{(1/2)}_{\text{II}})}{4G_N^{(3)}}+\frac{\mathcal{A}(C^{(1/2)}_{\text{III}})}{4G_N^{(3)}},
\end{equation}
\begin{equation}
S_{\text{gen}}^{(1/2)}(A\cup B)=\frac{\mathcal{A}^{(1/2)}(a)}{4G_N^{(2)}}+\frac{\mathcal{A}^{(1/2)}(a'')}{4G_N^{(2)}}+\frac{\mathcal{A}(C^{(1/2)}_{\text{I}})}{4G_N^{(3)}}+\frac{\mathcal{A}(C^{(1/2)}_{\text{III}})}{4G_N^{(3)}},
\end{equation}
the generalized entanglement negativity for phase-II from proposal \eqref{p2}  is 
\begin{equation}\label{pp2}
\begin{split}
    \mathcal{E}_{\mathrm{gen}}(A: B)
    =&\frac{\mathcal{A}^{1/2}(a')}{4G_N^{(2)}}+\frac{\mathcal{A}(C^{(1/2)}_{\text{II}})}{4G_N^{(3)}}.
    \end{split}
\end{equation}

Obviously, after minimization, eqs. \eqref{Egen_II1}, \eqref{ppp2} and \eqref{pp2} give the same results.

\subsection{Phase-III}
Consider the phase-III (Fig.\ref{Fig:21}) where the interval $B$ is too small to admit an island.
Firstly, according to \eqref{HS_negativity}, the effective entanglement negativity is 
\begin{equation}
    \mathcal{E}^{\text{eff}}=\frac{1}{2}\mathcal{A}(C^{(1/2)})_{\text{3-point}}=\frac{1}{2}\left(\frac{\mathcal{A}(C^{(1/2)}_{\text{II}})}{4G_N^{(3)}}+\frac{\mathcal{A}(C^{(1/2)}_{\text{IV}})}{4G_N^{(3)}}-\frac{\mathcal{A}(C^{(1/2)}_{\text{III}})}{4G_N^{(3)}}\right),
\end{equation}
thus  the generalized entanglement negativity from proposal \eqref{p3} is
\begin{equation}\label{phase3-1}
\begin{split}
    \mathcal{E}_{\mathrm{gen}}(A: B)=\mathcal{E}^{\text{eff}}
    =\frac{1}{2}\left(\frac{\mathcal{A}(C^{(1/2)}_{\text{II}})}{4G_N^{(3)}}+\frac{\mathcal{A}(C^{(1/2)}_{\text{IV}})}{4G_N^{(3)}}-\frac{\mathcal{A}(C^{(1/2)}_{\text{III}})}{4G_N^{(3)}}\right).
    \end{split}
\end{equation}

Secondly, the entanglement negativity from proposal eq.\eqref{p1} is given by half the minimal $S_{R~\rm gen}^{(1/2)}$.
In this case, $S_{R~\rm gen}^{(1/2)}$ is derived from the holographic scheme eq.\eqref{SR_eff} as 
\begin{equation}\label{phase3-2}
\begin{split}
   \frac12 S_{R~\rm gen}^{(1/2)}= \frac12 S^{(1/2,1)\text{eff}}_R=&
   \frac12\left( \mathcal{A}(C^{(1/2)})_{\text{2-point}}+\mathcal{A}(C^{(1/2)})_{\text{3-point}}\right)\\
    =&\frac12 \mathcal{A}(C^{(1/2)})_{\text{3-point}}\\
    =&\frac12\left(\frac{\mathcal{A}(C^{(1/2)}_{\text{II}})}{4G_N^{(3)}}+\frac{\mathcal{A}(C^{(1/2)}_{\text{IV}})}{4G_N^{(3)}}-\frac{\mathcal{A}(C^{(1/2)}_{\text{III}})}{4G_N^{(3)}}\right).
    \end{split}
\end{equation}

Thirdly, since we have
\begin{equation}
S_{\text{gen}}^{(1/2)}(A)=\frac{\mathcal{A}^{(1/2)}(a)}{4G_N^{(2)}}+\frac{\mathcal{A}^{(1/2)}(a')}{4G_N^{(2)}}+\frac{\mathcal{A}(C^{(1/2)}_{\text{I}})}{4G_N^{(3)}}+\frac{\mathcal{A}(C^{(1/2)}_{\text{II}})}{4G_N^{(3)}},
\end{equation}
\begin{equation}
S_{\text{gen}}^{(1/2)}(B)=\frac{\mathcal{A}(C^{(1/2)}_{\text{IV}})}{4G_N^{(3)}},
\end{equation}
\begin{equation}
S_{\text{gen}}^{(1/2)}(A\cup B)=\frac{\mathcal{A}^{(1/2)}(a)}{4G_N^{(2)}}+\frac{\mathcal{A}^{(1/2)}(a')}{4G_N^{(2)}}+\frac{\mathcal{A}(C^{(1/2)}_{\text{I}})}{4G_N^{(3)}}+\frac{\mathcal{A}(C^{(1/2)}_{\text{III}})}{4G_N^{(3)}},
\end{equation}
the generalized entanglement negativity for phase-III from proposal \eqref{p2}  is
\begin{equation}
\begin{split}\label{phase3-3}
    \mathcal{E}_{\mathrm{gen}}(A: B)=&\frac{1}{2}\left(\frac{\mathcal{A}(C^{(1/2)}_{\text{I}})}{4G_N^{(3)}}+\frac{\mathcal{A}(C^{(1/2)}_{\text{II}})}{4G_N^{(3)}}+\frac{\mathcal{A}(C^{(1/2)}_{\text{IV}})}{4G_N^{(3)}}-\frac{\mathcal{A}(C^{(1/2)}_{\text{I}})}{4G_N^{(3)}}-\frac{\mathcal{A}(C^{(1/2)}_{\text{III}})}{4G_N^{(3)}}\right)\\
    =&\frac{1}{2}\left(\frac{\mathcal{A}(C^{(1/2)}_{\text{II}})}{4G_N^{(3)}}+\frac{\mathcal{A}(C^{(1/2)}_{\text{IV}})}{4G_N^{(3)}}-\frac{\mathcal{A}(C^{(1/2)}_{\text{III}})}{4G_N^{(3)}}\right).
    \end{split}
\end{equation}

After minimization, eqs. \eqref{phase3-1}, \eqref{phase3-2} and \eqref{phase3-3} give the same results.

\subsection{Phase-IV}

For phase-IV (Fig.\ref{Fig:3}) where the sandwich interval $C$ admits an island, 
the entanglement negativity given in eq.\eqref{HS_negativity} is 
\begin{equation}
    \mathcal E(A:B)=0,
\end{equation}
which means there is no entanglement between $A$ and $B$ if the sandwich interval $C$ has an island.
The EWCS also vanishes.
The same is true for Renyi reflected entropy
\begin{equation}
\begin{split}
    S^{(1/2,1)\text{eff}}_R=& \mathcal{A}(C^{(1/2)})_{\text{2-point}}+\mathcal{A}(C^{(1/2)})_{\text{3-point}}=0.
    \end{split}
\end{equation}

Proposal eq.\eqref{p2} states that $\mathcal E(A:B)$ is given by the combination of the following generalized entanglement entropies
\begin{equation}
S_{\text{gen}}^{(1/2)}(A\cup C)=\frac{\mathcal{A}^{(1/2)}(a)}{4G_N^{(2)}}+\frac{\mathcal{A}^{(1/2)}(a'')}{4G_N^{(2)}}+\frac{\mathcal{A}(C^{(1/2)}_{\text{I}})}{4G_N^{(3)}}+\frac{\mathcal{A}(C^{(1/2)}_{\text{III}})}{4G_N^{(3)}},
\end{equation}
\begin{equation}
S_{\text{gen}}^{(1/2)}(B\cup C)=\frac{\mathcal{A}^{(1/2)}(a')}{4G_N^{(2)}}+\frac{\mathcal{A}^{(1/2)}(a''')}{4G_N^{(2)}}+\frac{\mathcal{A}(C^{(1/2)}_{\text{II}})}{4G_N^{(3)}}+\frac{\mathcal{A}(C^{(1/2)}_{\text{IV}})}{4G_N^{(3)}},
\end{equation}
\begin{equation}
S_{\text{gen}}^{(1/2)}(A\cup B\cup C)=\frac{\mathcal{A}^{(1/2)}(a)}{4G_N^{(2)}}+\frac{\mathcal{A}^{(1/2)}(a''')}{4G_N^{(2)}}+\frac{\mathcal{A}(C^{(1/2)}_{\text{I}})}{4G_N^{(3)}}+\frac{\mathcal{A}(C^{(1/2)}_{\text{IV}})}{4G_N^{(3)}},
\end{equation}
\begin{equation}
S_{\text{gen}}^{(1/2)}(C)=\frac{\mathcal{A}^{(1/2)}(a')}{4G_N^{(2)}}+\frac{\mathcal{A}^{(1/2)}(a'')}{4G_N^{(2)}}+\frac{\mathcal{A}(C^{(1/2)}_{\text{II}})}{4G_N^{(3)}}+\frac{\mathcal{A}(C^{(1/2)}_{\text{III}})}{4G_N^{(3)}}.
\end{equation}
Then the generalized entanglement negativity for phase-IV is 
\begin{equation}
\begin{split}
    \mathcal{E}_{\mathrm{gen}}(A: B)=&\frac{1}{2}\left(\frac{\mathcal{A}(C^{(1/2)}_{\text{I}})}{4G_N^{(3)}}+\frac{\mathcal{A}(C^{(1/2)}_{\text{III}})}{4G_N^{(3)}}+\frac{\mathcal{A}(C^{(1/2)}_{\text{II}})}{4G_N^{(3)}}+\frac{\mathcal{A}(C^{(1/2)}_{\text{IV}})}{4G_N^{(3)}}\right.\\
   &\left. -\frac{\mathcal{A}(C^{(1/2)}_{\text{I}})}{4G_N^{(3)}}-\frac{\mathcal{A}(C^{(1/2)}_{\text{IV}})}{4G_N^{(3)}}-\frac{\mathcal{A}(C^{(1/2)}_{\text{II}})}{4G_N^{(3)}}-\frac{\mathcal{A}(C^{(1/2)}_{\text{III}})}{4G_N^{(3)}}\right)\\
    =&0,
    \end{split}
\end{equation}
which agrees with the former proposals.

Above calculation in the four phases shows us the exact match of entanglement negativity from the three proposals eqs. \eqref{p3}, \eqref{p1} and \eqref{p2}. 
As a corollary, we conclude that the island 
$$ 
I_R=I_{\mathcal E}
$$
for general 2D eternal black holes coupled to CFT matters.

\section{Consistency check of island formula of reflected entropy for general 2D eternal black holes}\label{sec5}
\begin{figure}[htp]
\centering
\includegraphics[width=0.8\textwidth]{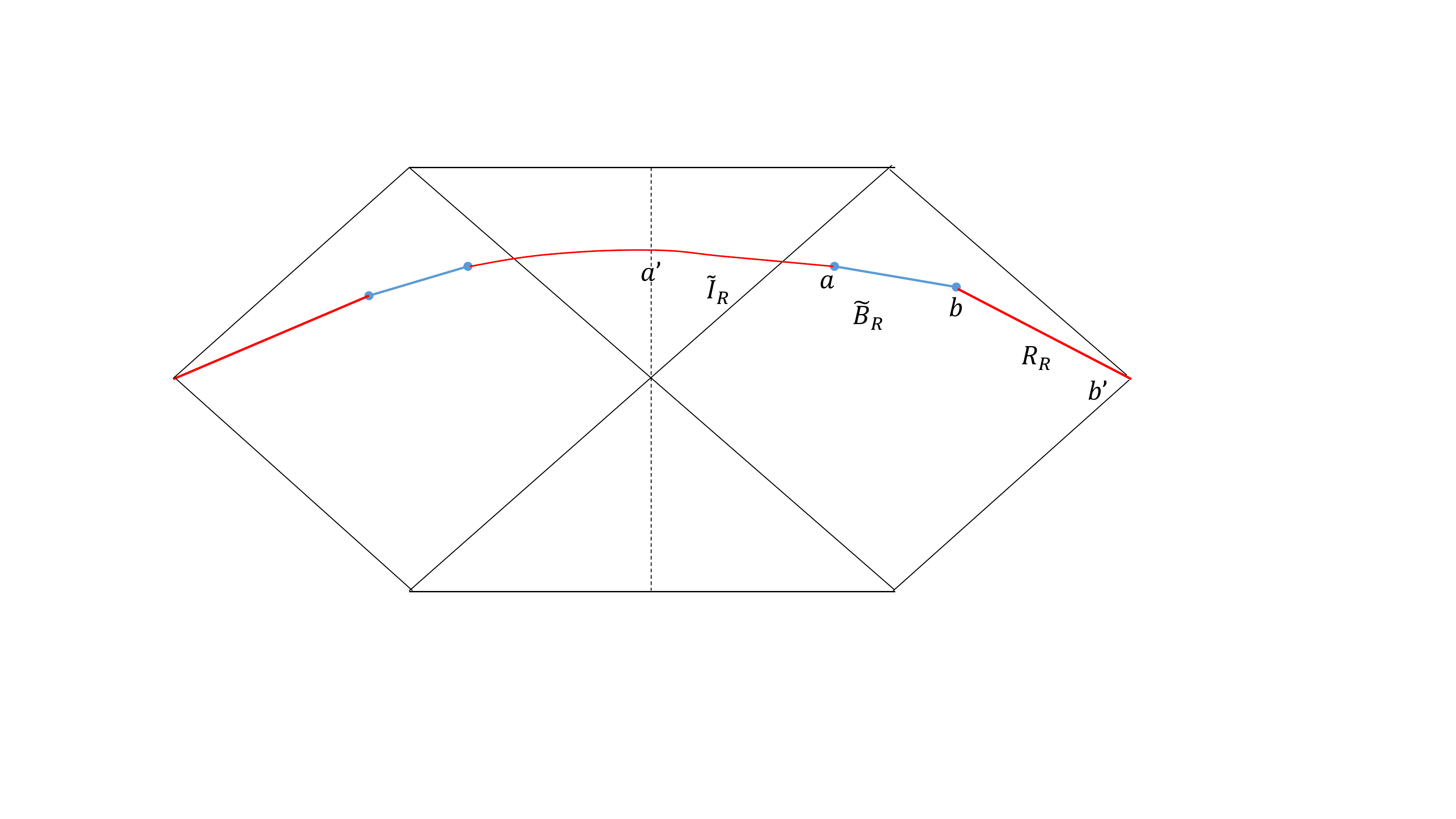}
\includegraphics[width=0.8\textwidth]{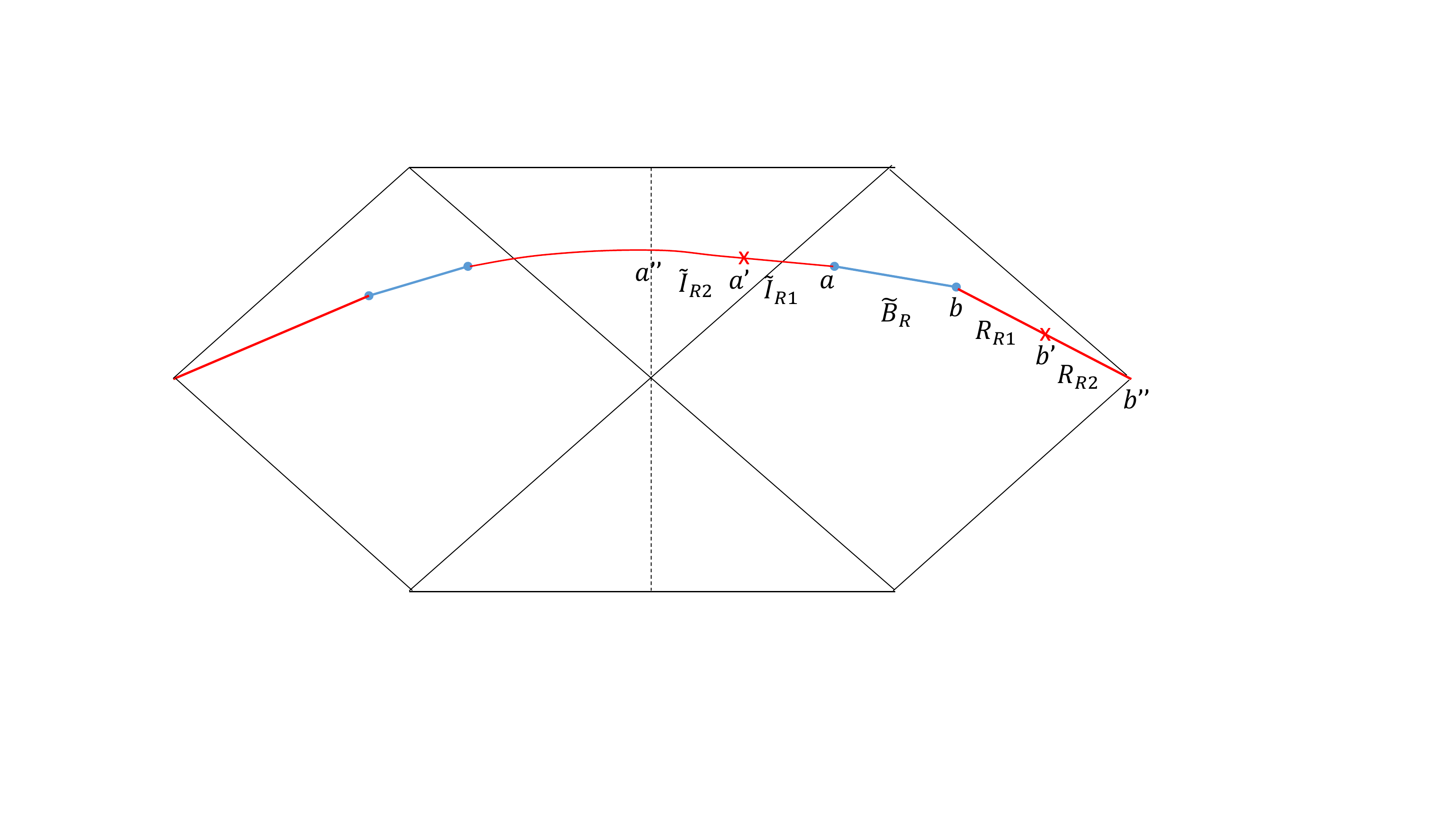}
\caption{Penrose diagram for general 2D eternal black holes. }
\label{Fig:4}
\end{figure}

In this section, we will use our holographic scheme for effective Renyi reflected entropy to study reflected entropy for general 2D eternal black holes coupled to CFT matters (Fig. \ref{Fig:4}). 
Note that for AdS black holes, to make black holes evaporate, we glue the original AdS spacetime with a flat spacetime along the boundary and impose the transparent boundary condition.
When coupled to semi-classical gravity, the island formula for reflected entropy is given by \cite{Chandrasekaran:2020qtn} 
\begin{equation}
\begin{aligned}
S_{R}(A: B)=\frac{2\operatorname{Area}\left[Q''=\partial \mathrm{I}_{R}(A) \cap \partial \mathrm{I}_{R}(B)\right]}{4G_{N}}+S_{R}^{(1)\mathrm{eff}}\left(A \cup I_{R}(A): B \cup I_{R}(B)\right).
\end{aligned}
\end{equation}
Note that a proper formula for reflected entropy should satisfy the following inequality \cite{Dutta:2019gen}
\begin{equation}\label{ineq}
I(A: B) \leq S_{R}(A: B) \leq 2 \min \{S(A), S(B)\},
\end{equation}
where 
\begin{equation}
    I(A:B)=S(A)+S(B)-S(A\cup B)
\end{equation}
is the mutual information between $A$ and $B$.
In this section, we will study this inequality for general 2D eternal black holes utilizing the holographic scheme developed in Sec.\ref{sec2}.

\subsection{Black holes and radiation}
\begin{figure}[htp]
\centering
\includegraphics[width=0.8\textwidth]{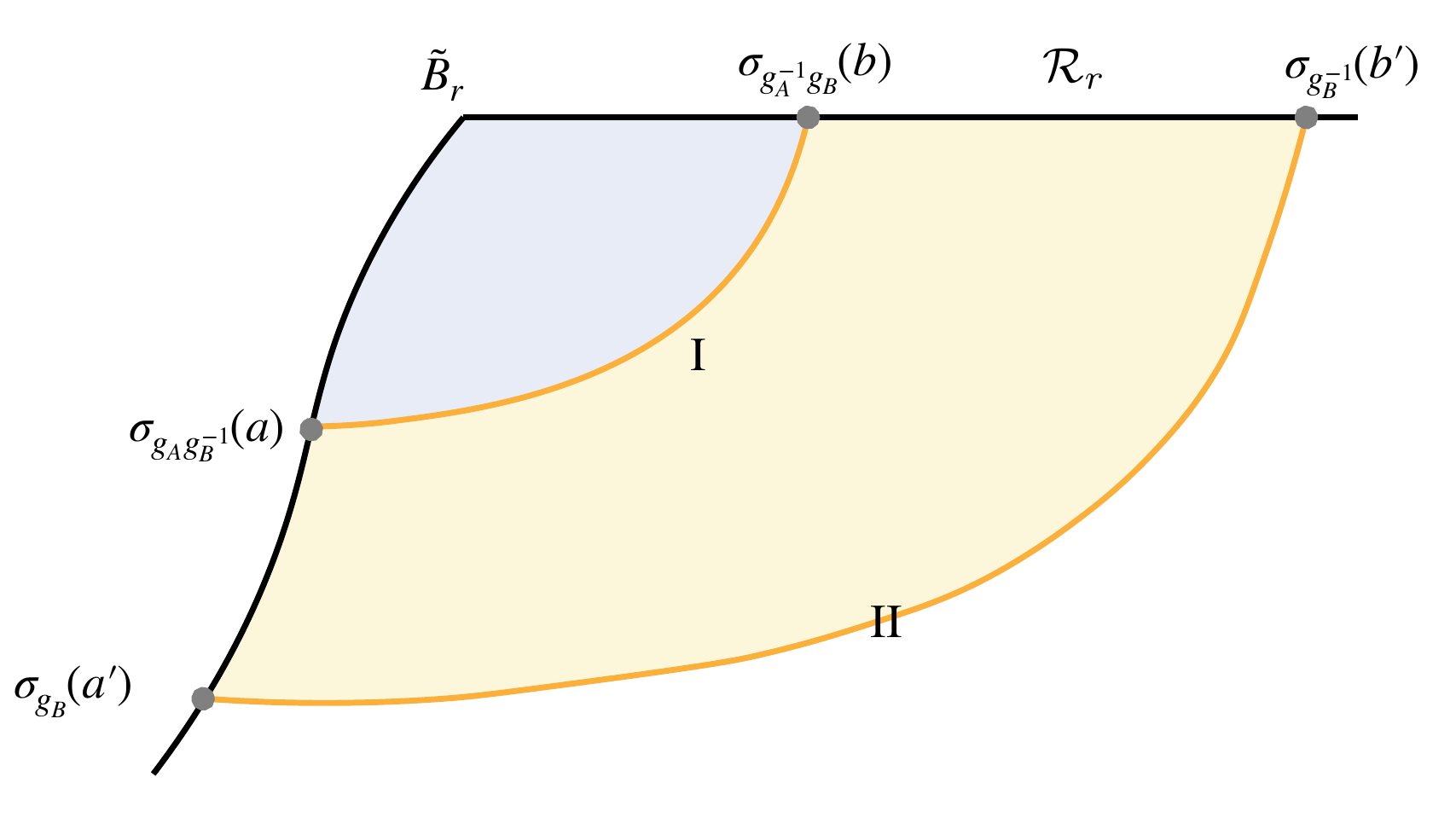}
\caption{Reflected entropy between black holes and radiation.
The light-blue and light-yellow region are the entanglement wedge of black holes and radiation, respectively. }
\label{Fig:5}
\end{figure}

We first study reflected entropy between the right black hole $\tilde B _{r}$ and right radiation $\mathcal R_r$ (Figs. \ref{Fig:4} and \ref{Fig:5}). 
Using the holographic scheme in Sec. \ref{sec2}, we get the effective reflected entropy
\begin{equation}
    S^{\text{eff}}_R(\tilde B _r: \mathcal R_r\cup \tilde{I}_\mathcal{R})= \mathcal{A}(C^{(1)})_{\text{2-point}}+\mathcal{A}(C^{(1)})_{\text{3-point}}=2\frac{\mathcal{A}(C^{(1)}_{ab})}{4G_N^{(3)}},
\end{equation}
which is nothing but the effective entanglement entropy between the two-sided black holes and radiations.
In this case, QEWCS and QES coincide, and we have 
\begin{equation}
    S_R(\tilde B _r: \mathcal R_r)=S(\tilde B:\mathcal R)=2\frac{\text{Area}(a)}{4G_N^{(2)}}+2\frac{\mathcal{A}(C^{(1)}_{ab})}{4G_N^{(3)}}.
\end{equation}
Now let us check the inequality \eqref{ineq}.
The entanglement entropy
\begin{equation}
\begin{split}
   &S(\tilde B _r)=\frac{\text{Area}(a)}{4G_N^{(2)}}+\frac{\mathcal{A}(C^{(1)}_{ab})}{4G_N^{(3)}},\\
   &S(\mathcal R_r)=\frac{\text{Area}(a)}{4G_N^{(2)}}+\frac{\text{Area}(a')}{4G_N^{(2)}}+\frac{\mathcal{A}(C^{(1)}_{ab})}{4G_N^{(3)}}+\frac{\mathcal{A}(C^{(1)}_{a'b'})}{4G_N^{(3)}},\\
    &S(\tilde B _r\cup \mathcal R_r )=\frac{\text{Area}(a')}{4G_N^{(2)}}+\frac{\mathcal{A}(C^{(1)}_{a'b'})}{4G_N^{(3)}}.
    \end{split}
\end{equation}
Thus the mutual information is
\begin{equation}
\begin{split}
    I(\tilde B _r: \mathcal R_r )    =&S(\tilde B _r)+S(\mathcal R_r)-S(\tilde B _r\cup \mathcal R_r )\\
    =&2\frac{\text{Area}(a)}{4G_N^{(2)}}+2\frac{\mathcal{A}(C^{(1)}_{ab})}{4G_N^{(3)}}\\
    =&S_R(\tilde B _r: \mathcal R_r ).
    \end{split}
\end{equation}
and
\begin{equation}
    \begin{split}
        2~\text{min}\{S(\tilde B _r),S(\mathcal R_r)\}=2S(\tilde B _r)
        =2\frac{\text{Area}(a)}{4G_N^{(2)}}+2\frac{\mathcal{A}(C^{(1)}_{ab})}{4G_N^{(3)}}=S_R(\tilde B _r: \mathcal R_r ).
    \end{split}
\end{equation}
In fact, in this case, the Araki-Lieb inequality is saturated
$$
S(\mathcal R_r)-S(\tilde B _r)=S(\mathcal R_r\cup \tilde B _r),
$$
thus we have \cite{Nguyen:2017yqw}
\begin{equation}
    \begin{split}
    & I(\tilde B _r: \mathcal R_r )=2S(\tilde B _r)\ 
    \Longrightarrow    I(\tilde B _r: \mathcal R_r )=S_R(\tilde B _r: \mathcal R_r )=2S(\tilde B _r)= 2\text{min}\{S(\tilde B _r),S(\mathcal R_r)\} ,
    \end{split}
\end{equation}
that is, the reflected entropy (RE) coincides with mutual information (MI) in this case.
In fact, RE=MI seems to be universal for 3D Chern-Simons theories \cite{Berthiere:2020ihq}.
It was also shown in Ref. \cite{Zou:2020bly} that 
for a pure state $|\psi\rangle_{ABC}$ in Hilbert space $\mathcal H_{ABC}$, 
if the Hilbert space can be factorized and the mutual information is equal to reflected entropy, i.e.
$$
h(A:B)=S_R(A:B)-I(A:B)=0,
$$
the state  $|\psi\rangle_{ABC}$ is in the sum of triangle state (SOTS), i.e.
\begin{equation}
|\psi\rangle_{A B C}=\sum_{j} \sqrt{p_{j}}|\psi_j\rangle_{A B C},
\end{equation}
where
\begin{equation}
|\psi_j\rangle_{A B C}=\left|\psi_{j}\right\rangle_{A_{R}^{j} B_{L}^{j}}\left|\psi_{j}\right\rangle_{B_{R}^{j} C_{L}^{j}}\left|\psi_{j}\right\rangle_{C_{R}^{j} A_{L}^{j}}
\end{equation}
is the triangle state without nontrivial tripartite entanglement.
However, here we comment that although MI=RE between $\mathcal R_r$ and $\tilde B _r$, whether the whole spacetime is in SOTS or not remains an open question due to not-well-defined factorization of Hilbert space involving gravity \cite{Donnelly:2016auv,Raju:2021lwh}. 

\subsection{Radiation and radiation}

\begin{figure}[htp]
\centering
\includegraphics[width=0.8\textwidth]{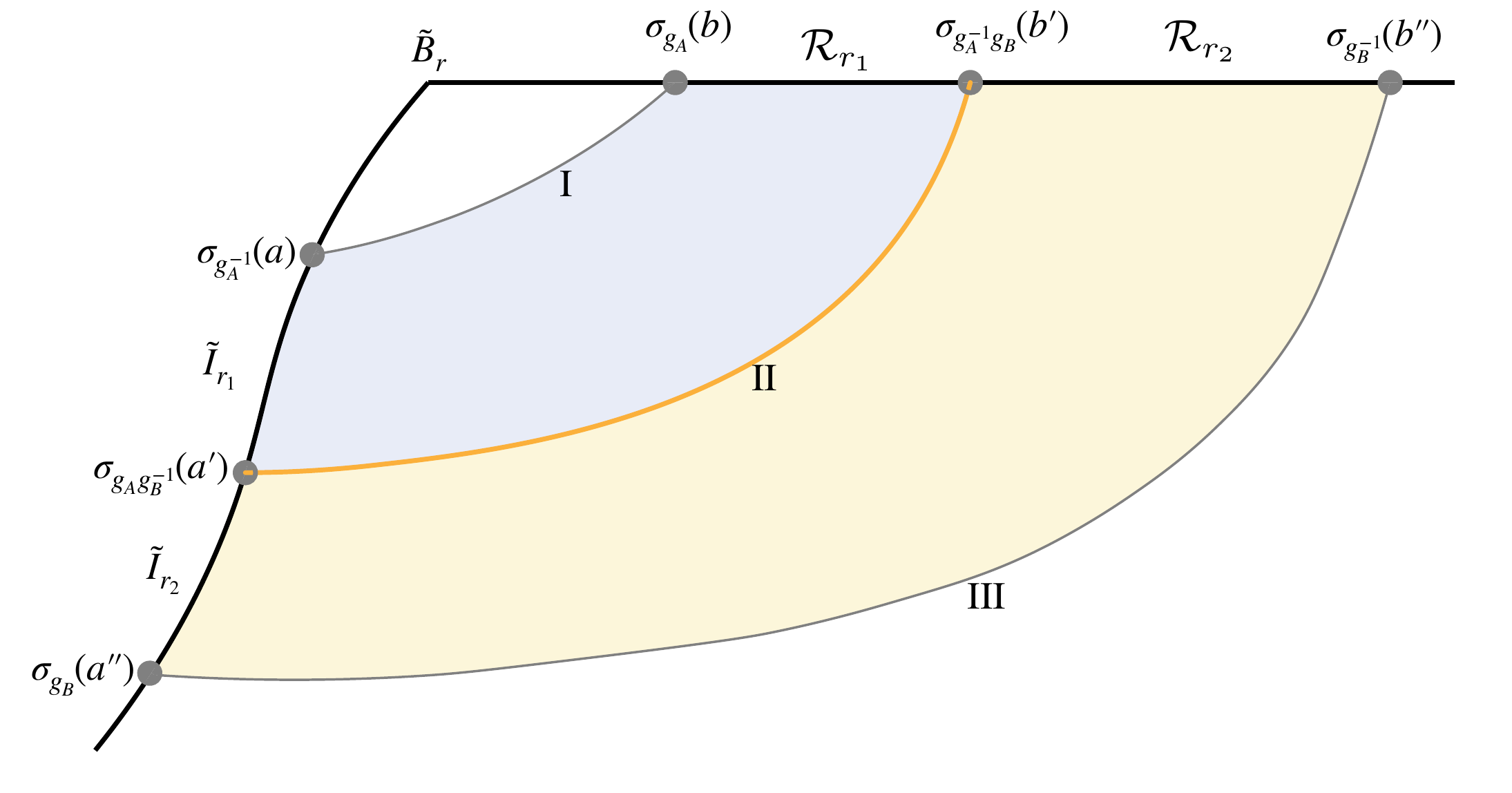}
\caption{Reflected entropy between two radiation regions. 
The light-blue and light-yellow region show the entanglement wedge of radiation $\mathcal R_{r_1}$ and radiation $\mathcal R_{r_2}$, respectively.}
\label{Fig:6}
\end{figure}

Now we study the reflected entropy between two radiation regions (Fig. \ref{Fig:6}). 
We first consider the case in which both radiation intervals $\mathcal R_{r_1}$ and $\mathcal R_{r_2}$ are large enough to admit islands. 
Using the holographic scheme in Sec.\ref{sec2}, we get the effective reflected entropy 
\begin{equation}
    S^{\text{eff}}_R(\mathcal R_{r_1}\cup \tilde{I}_{r_1}: \mathcal R_{r_2}\cup \tilde{I}_{r_2} )= \mathcal{A}(C^{(1)})_{\text{2-point}}+\mathcal{A}(C^{(1)})_{\text{3-point}}=2\frac{\mathcal{A}(C^{(1)}_{a'b'})}{4G_N^{(3)}}.
\end{equation}
Then the reflected entropy is given by
\begin{equation}
    S_R(\mathcal R_{r_1}: \mathcal R_{r_2} )= 2\frac{\text{Area}(a')}{4G_N^{(2)}}+ 2\frac{\mathcal{A}(C^{(1)}_{a'b'})}{4G_N^{(3)}}.
\end{equation}
Now let us check the inequality \eqref{ineq}.
The entanglement entropies for $\mathcal {R}_{r_1}$, $\mathcal {R}_{r_2}$ and their union $\mathcal R_{r_1}\cup \mathcal R_{r_2}$ are given by
\begin{equation}
\begin{split}
   &S(\mathcal R_{r_1})=\frac{\text{Area}(a)}{4G_N^{(2)}}+\frac{\text{Area}(a')}{4G_N^{(2)}}+\frac{\mathcal{A}(C^{(1)}_{ab})}{4G_N^{(3)}}+\frac{\mathcal{A}(C^{(1)}_{a'b'})}{4G_N^{(3)}},\\
   &S(\mathcal R_{r_2})=\frac{\text{Area}(a')}{4G_N^{(2)}}+\frac{\text{Area}(a'')}{4G_N^{(2)}}+\frac{\mathcal{A}(C^{(1)}_{a'b'})}{4G_N^{(3)}}+\frac{\mathcal{A}(C^{(1)}_{a''b''})}{4G_N^{(3)}},\\
    &S(\mathcal R_{r_1}\cup \mathcal R_{r_2} )=\frac{\text{Area}(a)}{4G_N^{(2)}}+\frac{\text{Area}(a'')}{4G_N^{(2)}}+\frac{\mathcal{A}(C^{(1)}_{ab})}{4G_N^{(3)}}+\frac{\mathcal{A}(C^{(1)}_{a''b''})}{4G_N^{(3)}},
    \end{split}
\end{equation}
thus the mutual information is
\begin{equation}
\begin{split}
    I(\mathcal R_{r_1}: \mathcal R_{r_2} )    =&S(\mathcal R_{r_1})+S(\mathcal R_{r_2})-S(\mathcal R_{r_1}\cup \mathcal R_{r_2} )\\
    =&2\frac{\text{Area}(a')}{4G_N^{(2)}}+2\frac{\mathcal{A}(C^{(1)}_{a'b'})}{4G_N^{(3)}}\\
    =&S_R(\mathcal R_{r_1}: \mathcal R_{r_2}  ).
    \end{split}
\end{equation}
and
\begin{equation}
    \begin{split}
        2\text{min}\{S(R_{R_1}),S(R_{R_2})\}=&2S(R_{R_1})\\
        =&2\frac{\text{Area}(a)}{4G_N^{(2)}}+2\frac{\text{Area}(a')}{4G_N^{(2)}}+2\frac{\mathcal{A}(C^{(1)}_{ab})}{4G_N^{(3)}}+2\frac{\mathcal{A}(C^{(1)}_{a'b'})}{4G_N^{(3)}}\\
        >&2\frac{\text{Area}(a')}{4G_N^{(2)}}+2\frac{\mathcal{A}(C^{(1)}_{a'b'})}{4G_N^{(3)}}=S_R(\mathcal R_{r_1}:\mathcal  R_{r_2}  ).
    \end{split}
\end{equation}
The saturation of this case is an analog of GHZ state \cite{Rangamani:2015qwa}
\begin{equation}
|\mathrm{GHZ}\rangle=\frac{1}{\sqrt{2}}(|0_A0_B0_C\rangle+|1_A1_B1_C\rangle),
\end{equation}
where the reflected entropy $S_R(A:B)$ only saturates the lower bound of the inequality \eqref{ineq}.

\section{Conclusion}\label{sec7}
By performing the replica trick to calculate the effective Renyi reflected entropy and entanglement negativity in large $c$ limit, we extract a holographic calculation scheme for effective Renyi reflected entropy and effective entanglement negativity for general 2D eternal black holes. 
We find that both effective Renyi reflected entropy and entanglement negativity can be expressed in terms of the combination of the areas of modified backreacting cosmic branes in $\text{AdS}_3$ bulk. 

For effective Renyi reflected entropy, the terms involving the scaling dimensions $h_{g_A},h_m$ will be cancelled out, which means the modified backreacting cosmic branes connecting island side and radiation side, where two twist operators $\sigma_{h_{g_A}}$  and  $\sigma_{h_{g_A^{-1}}}$ inserted,  do not contribute. 
The contribution is from the cosmic branes that connect two twist operators with scaling dimension $h_{g_Ag_B^{-1}}$ and $h_{g_Bg_A^{-1}}$, and that connect three twist operators with scaling dimension $h_{g_Ag_B^{-1}}$, $h_{g_A^{-1}}$ and $h_{g_B}$.

For effective entanglement negativity, we find that the scaling dimension $h_{\tau_{n_{e}}}$ of the twist operator $\tau_{n_e}$ becomes zero as Renyi index $n_e\to1$.
Therefore, the effective entanglement negativity receives no contribution from the modified backreacting cosmic branes connecting the two twist operators $\tau_{n_e},\bar{\tau}_{n_e}$.
However, the cosmic branes that connect two twist operators with scaling dimensions $h_{\tau^2_{n_{e}}},h_{\bar{\tau}^2_{n_{e}}}$, and that connect three twist operators with scaling dimensions $h_{\tau_{n_e}},h_{\tau_{n_e}},h_{\bar{\tau}^2_{n_e}}$ do contribute.

With the holographic scheme in hand, we then investigate the consistency condition for island formulae for entanglement negativity and reflected entropy. 

There are two proposals to calculate the entanglement negativity.
One is given by minimal QEWCS proposal.
The other is given by a combination of Renyi entropy of 1/2 order when two intervals $A$ and $B$ are in proximity. 
They provide a consistency check for the island formula of entanglement negativity. 
Using the holographic scheme, the two proposals give the same entanglement negativity for general 2D eternal black holes. 
As a corollary, we deduce that the islands for reflected entropy and entanglement negativity coincide with each other for general 2D eternal black holes coupled to CFT matters.

The reflected entropy inequality $I(A: B) \leq S_{R}(A: B) \leq 2 \min \{S(A), S(B)\}$ provides a consistency check for island formula for reflected entropy.
We study the saturation of this inequality for general 2D eternal black holes by utilizing the holographic calculation scheme developed in Sec. \ref{sec2}. 
We find that reflected entropy between the black hole and the radiation saturates the inequality due to the saturation of  Araki-Lieb inequality. 
For reflected entropy between two radiation regions, we find that reflected entropy only saturates the lower bound of the inequality, which is an analog of GHZ state.

\begin{acknowledgments}
We thank Liang Ma for useful discussions.
The authors were supported in part by the National Natural Science Foundation of China under Grant No. 11875136 and the Major Program of the National Natural Science Foundation of China under Grant No. 11690021. 
\end{acknowledgments}

%

\end{document}